\PassOptionsToPackage{sort&compress}{natbib}
\documentclass{arxivA} 

\usepackage{amsmath,amssymb,amsfonts,amsthm,makeidx,graphicx}

\usepackage{hyperref}
\usepackage{slashed}
\usepackage[table]{xcolor}
\usepackage{booktabs}
\usepackage{makecell}

\definecolor{gluonred}{RGB}{125,55,70}
\definecolor{gluonlight}{RGB}{245,230,233}
\definecolor{gluonrow}{RGB}{250,245,246}
\definecolor{tmdteal}{RGB}{30,90,100}
\definecolor{tmdlight}{RGB}{224,239,239}
\definecolor{tmdrow}{RGB}{243,248,248}
\definecolor{nsgold}{RGB}{153,105,35}
\definecolor{nsgoldlight}{RGB}{247,239,220}
\definecolor{nsgoldrow}{RGB}{251,248,240}
\definecolor{mvblue}{RGB}{55,75,110}
\definecolor{mvbluelight}{RGB}{226,233,243}
\definecolor{mvbluerow}{RGB}{243,246,250}

\newcommand{\der}{\textrm{d}}
\newcommand{\Pt}{\boldsymbol{P}_\perp}

\newcommand{\kt}{\boldsymbol{k}_\perp}
\newcommand{\ellt}{\boldsymbol{\ell}_\perp}

\begin{document}

\chapter{Transverse momentum dependent\\ parton distributions at small $x$}\label{chap1}

\author[1]{Paul Caucal}
\address[1]{SUBATECH UMR 6457 (IMT Atlantique, Université de Nantes, IN2P3/CNRS), 4 rue Alfred Kastler, 44307 Nantes, France} 
\author[2,3,4]{Farid Salazar}
\address[2]{Department of Physics, Temple University, Philadelphia, PA 19122 - 1801, USA}
 \address[3]{RIKEN-BNL Research Center, Brookhaven National Laboratory, Upton, New York 11973, USA}
 \address[4]{Physics Department, Brookhaven National Laboratory, Upton, New York 11973, USA}

\maketitle

\begin{abstract}[Abstract]

Transverse-momentum-dependent (TMD) parton distributions provide a three-dimensional description of the momentum structure of hadrons. They are probed in proton--proton and proton--nucleus collisions at the LHC and RHIC, and are a central part of the physics program of the future Electron-Ion Collider (EIC). While TMD factorization is well established at moderate Bjorken $x$, the small-$x$ regime requires a different treatment, since large logarithms of $1/x$ must be resummed and non-linear effects due to high gluon densities become important. The Color Glass Condensate (CGC) effective theory provides a natural framework to describe TMDs in this saturation regime and to connect them with standard TMD factorization.

This chapter reviews quark and gluon TMDs at small $x$. We discuss operator definitions, gauge-link structures and representation in terms of Wilson-line correlators in the CGC, for unpolarized and polarized distributions, including linearly polarized gluons and the spin-dependent odderon. We review the factorization formulae for inclusive and diffractive observables, highlighting the role of process dependence and the emergence of a limited set of Wilson-line operators. We then summarize the evolution of these distributions, namely the resummation of logarithms of $1/x$ by the Balitsky--Kovchegov and JIMWLK equations, and of Sudakov logarithms of the hard scale through Collins--Soper--Sterman evolution. We also review TMDs beyond the eikonal approximation, including the sub-eikonal Wilson-line operators that give access to helicity and other spin-dependent distributions, their double-logarithmic small-$x$ evolution and the resulting asymptotics, and the sub-eikonal corrections to the eikonal distributions, which provide a first step toward the moderate-$x$ regime. Finally, we discuss phenomenological implications for present and future experiments, with emphasis on the EIC, and outline the theoretical challenges toward a unified description of TMDs.

\end{abstract}

\begin{keywords}
High-energy Quantum Chromodynamics, small-$x$ physics, gluon saturation, transverse-momentum-dependent parton distributions, Color Glass Condensate, Electron-Ion Collider
\end{keywords}

\newpage

\makeatletter
\renewcommand{\tableofcontents}{%
  \@starttoc{toc}%
}
\makeatother

{\small\setstretch{1.0}\setcounter{tocdepth}{2}\tableofcontents}
\bigskip

\section{Introduction}

The transverse-momentum-dependent (TMD) factorization framework has become one of the central theoretical tools for studying the three-dimensional momentum structure of hadrons~\cite{Collins:1981uk,Collins:1981uw,Collins:1984kg,Ji:2004wu,Collins:2011zzd,Boussarie:2023izj}. By separating the perturbatively calculable short-distance dynamics from the universal non-perturbative TMD parton distribution functions (TMD PDFs), it provides a window to connect experimental measurements to the intrinsic transverse motion of quarks and gluons inside the proton. The resummation of Sudakov logarithms and the treatment of rapidity divergences at moderate Bjorken $x$ have reached a level of maturity that supports increasingly precise extractions of proton structure (see~\cite{Boussarie:2023izj} for a recent review).

However, the small-$x$ regime presents a qualitatively different dynamical situation. As in the case of collinear factorization, the perturbative expansion underlying the standard TMD formalism becomes unstable through the appearance of large logarithms of $1/x$, which must be resummed to all orders by the Balitsky--Fadin--Kuraev--Lipatov (BFKL) equation~\cite{Kuraev:1977fs,Balitsky:1978ic}, as also suggested by collinear fits to HERA data~\cite{Caola:2009iy,Ball:2017otu,xFitterDevelopersTeam:2018hym}. Without such a resummation the perturbative coefficients may cease to converge in a controlled way, and the framework developed for moderate $x$ cannot simply be extrapolated downward in $x$. Addressing these challenges is a key goal of this chapter.

A second motivation for the present chapter arises from nuclear physics. While TMD factorization has been developed extensively for proton targets, its extension to nuclei remains considerably less understood, since the multiple scattering, coherent interactions and high gluon densities characteristic of nuclear targets introduce dynamical effects that lie outside the standard moderate-$x$ formalism~\cite{Gribov:1984tu,Mueller:1985wy,McLerran:1993ni,McLerran:1993ka}. A first-principles understanding of TMDs in nuclei is therefore one ingredient of the physics program of the future Electron-Ion Collider (EIC)~\cite{Accardi:2012qut,AbdulKhalek:2021gbh,AbdulKhalek:2022hcn}, whose scientific goals include imaging the three-dimensional gluon structure of both protons and nuclei in the high-energy regime, as emphasized in recent community assessments of the cold-nuclear and precision-QCD programs~\cite{Arleo:2025oos,Proceedings:2026xrb}.

A first-principles determination of these distributions, in protons and even more so in nuclei, requires non-perturbative methods, and lattice QCD is advancing steadily toward calculations of TMDs. Since the TMD operators involve Wilson lines extending to light-like infinity, they cannot be evaluated directly in a Euclidean path integral, and instead one has to evaluate quasi-distributions with finite space-like Wilson lines and match perturbatively onto the light-cone objects~\cite{Musch:2010ka,Ji:2013dva,Ebert:2019okf} (see also more recently Coulomb-gauge-dressed correlators that dispense with Wilson lines altogether~\cite{Zhao:2023ptv}). First-principles determinations of the Collins--Soper kernel exist~\cite{Shanahan:2020zxr}, and lattice calculations of parton structure have been surveyed recently~\cite{Constantinou:2020hdm}. While the present calculations are well suited to moderate $x$, reaching small $x$ requires correlators at large Ioffe time, where the signal-to-noise ratio degrades exponentially with the hadron momentum and with the length of the Wilson line. The matching onto light-cone distributions also requires the hadron momentum to be large compared with all other scales, which the lattice cutoff limits~\cite{Constantinou:2020hdm}. Nuclei pose a separate challenge: the number of Wick contractions grows factorially with the number of quarks~\cite{Detmold:2012eu,Doi:2012xd}, and the signal-to-noise ratio falls exponentially with the mass number~\cite{Lepage:1989hd}. Lattice calculations of nuclear matrix elements have so far been limited to nuclei with $A\leq3$, at unphysical pion masses, and to simple quantities such as charges, magnetic moments and the lowest moments of parton distributions~\cite{Beane:2014ora,Winter:2017bfs,Chang:2017eiq,Detmold:2020snb,Davoudi:2020ngi}. Furthermore, as we will see in the next section, the natural degrees of freedom at small $x$ are multi-point Wilson-line correlators, which are much more complex than the typical two- or three-point functions computed on the lattice.

These difficulties have a physical origin. At small $x$, the rapid growth of the gluon occupation numbers turns the hadronic wave function into a dense many-body system, in which the non-linear interactions are as important as the emissions responsible for the growth itself. In that regime lightlike Wilson lines are the fundamental degrees of freedom, which encode multiple scattering. For this reason the descriptions of TMDs at small $x$ that have proved most useful approach the problem directly from the high-energy limit $x\ll 1$, rather than extending the moderate-$x$ formalism downward. In this limit the Color Glass Condensate (CGC) effective theory~\cite{Iancu:2003xm,Gelis:2010nm,Kovchegov:2012mbw} and the high-energy operator product expansion formalism~\cite{Balitsky:1995ub} provide a common description both of the non-linear evolution of the hadronic wave function and of the transverse momentum distributions probed in scattering processes. Corrections away from the strict small-$x$ limit may then be incorporated systematically, which is how one might hope to understand the transition toward the conventional TMD factorization valid at moderate $x$.

Throughout this review, we will encounter multiple momentum scales. The center-of-mass energy $\sqrt{s}$ of the collision is much larger than any other scale in the problem. We will denote the hard scale in the observable as $Q$, and the intrinsic transverse momentum of the small-$x$ distribution as $k_\perp$. The saturation scale $Q_s(x)$ is an emergent energy- and nuclear-dependent transverse momentum scale which controls the transition between linear and non-linear evolution. The TMD regime at small $x$ is valid under the hierarchy:
\begin{align}
    \Lambda^2_{\mathrm{QCD}} \lesssim \left\{Q_s^2(x),\,k_\perp^2\right\} \ll Q^2 \ll s\,,
\end{align}
with no hierarchy assumed between $k_\perp$ and $Q_s$. The last inequality, $Q^2\ll s$, guarantees the small-$x$ regime, since the longitudinal momentum fraction probed in the target is parametrically $x\sim Q^2/s\ll1$, its precise value depending on the final-state kinematics, while $k_\perp^2\ll Q^2$ defines the TMD regime. Within this hierarchy we distinguish the dilute regime when $Q_s^2(x) \lesssim k_\perp^2$ and the dense regime $k_\perp^2 \lesssim Q_s^2(x)$.

The purpose of this chapter is to examine our current understanding of quark and gluon TMD distributions in the small-$x$ regime. Sec.~\ref{sec:tmds} covers the operator definitions, their gauge-link structures and the Wilson-line correlators onto which they map as $x\to0$, for the unpolarized, polarized, diffractive and target-fragmentation cases. In Sec.~\ref{sec:evolution}, we turn to quantum corrections, focusing in particular on the resummation of $\ln(1/x)$ logarithms through the Balitsky--Kovchegov (BK) and JIMWLK equations~\cite{Balitsky:1995ub, Kovchegov:1999yj,Iancu:2001ad, Jalilian-Marian:1997qno,Jalilian-Marian:1997ubg,Kovner:2000pt, Iancu:2001md,Ferreiro:2001qy}, as well as the resummation of Sudakov logarithms of $Q/k_\perp$, that appear at next-to-leading order. Sec.~\ref{sec:beyond-eikonal} goes beyond the eikonal approximation, which is required for the helicity and other spin-dependent distributions that vanish at leading power in $x$, and which provides a first step toward the moderate-$x$ regime. In Sec.~\ref{sec:pheno} we review the observables through which these distributions are accessed, focusing on the unpolarized case (for which experimental data are currently available in the small-$x$ regime), from semi-inclusive deep-inelastic scattering (SIDIS) and two-particle correlations to heavy flavor, diffraction and energy correlators, together with the comparisons with data available so far. Sec.~\ref{sec:advanced} collects topics that connect small-$x$ TMDs to neighboring areas, namely generalized TMDs and orbital angular momentum, power corrections in $k_\perp/Q$, Monte Carlo event generators and heavy-ion collisions. We summarize the main open problems in Sec.~\ref{sec:outlook}. 

\section{Gluon and quark TMDs at small $x$: operator definitions}
\label{sec:tmds}

In this section we introduce the gluon and quark TMDs that contribute at leading power (leading eikonal accuracy) at small $x$. Building upon the first review on this topic~\cite{Petreska:2018cbf}, we place particular emphasis on the connection between the gauge-link structure appearing in their operator definitions and the corresponding Wilson-line operators in the CGC framework. This correspondence allows their $x$ and $k_\perp$ dependence to be described in terms of saturation physics. We begin by making precise what is meant by factorization in this context, because in the small-$x$ literature it usually refers to fixed-order results for particular processes, whereas at moderate $x$ it refers to all-order theorems. We then divide the discussion into unpolarized and polarized TMDs, since only a subset of the quark and gluon TMDs remains leading power in $x$ in polarized processes at small $x$.

Two families of transverse-momentum-dependent objects appear in the high-energy literature. The \emph{unintegrated gluon distribution} (UGD) is defined through a factorization formula: a cross section computed at fixed order in $\alpha_s$ in the high-energy limit takes the form of an off-shell hard factor convoluted with a $k_\perp$-dependent function of the target, which is identified as the UGD~\cite{Gribov:1984tu,Catani:1990eg,Catani:1994sq}. The \emph{TMD} is defined as a matrix element of a bilocal operator with a specified gauge link, as in Eq.~\eqref{eq:gTMD-def}, independently of any process~\cite{Collins:1981uw,Collins:2011zzd,Bomhof:2006dp}. The former is tied to the order and process from which it was extracted, whereas the latter is a property of the target, and a factorization theorem asserts that the cross section depends on this operator alone to all orders, with the process dependence confined to the perturbative hard factor. The Weizs\"acker--Williams and dipole distributions introduced below can be understood in both ways: they arise as the $k_\perp$-dependent functions of the target in leading-order cross sections, and they also possess operator definitions in terms of Wilson lines. At leading order and in the correlation limit, where the resolved transverse momentum is small compared with the hard scale, the two definitions agree, and certain cross sections depend only on these correlators \cite{Dominguez:2010xd,Dominguez:2011wm}. Away from this limit the two need not agree, even at leading order: in the dilute region $k_\perp\gg Q_s$ the dependence on the gauge-link structure drops out and all leading-power gluon distributions, 	the Weizs\"acker--Williams (WW) and dipole among them, collapse to a single universal two-point function, which is why high-energy factorization (HEF) involves one UGD only, whereas in the saturation region the cross section involves higher-point Wilson-line correlators and no longer takes the form of one $k_\perp$-dependent function convoluted with an off-shell hard factor~\cite{Dominguez:2011wm,Kotko:2015ura}. We use the word factorization for these results in a deliberately limited sense: they show that a particular cross section factorizes at that order, and no all-order argument of the kind available at moderate $x$ exists for any of them. The distinction between the two definitions becomes operative at next-to-leading order (NLO), where the calculations of Sec.~\ref{sec:sudakov} test whether the same operators reappear, since a quantity that is universal at 
leading order may acquire process-dependent contributions at one loop.

At moderate $x$, standard TMD factorization is violated for processes in which color flows through both the initial and the final state~\cite{Collins:2007nk,Rogers:2010dm}: the Glauber exchanges cancel only when the associated color rotations commute, which fails with three or more colored external legs. The process-dependent gauge links of~\cite{Bomhof:2006dp} keep track of the color flow, but they make the TMDs process dependent.  At small $x$, the CGC treats the target to all orders in its color field, so that the multiple scattering left uncanceled at moderate $x$ is resummed into the Wilson-line operators, and at leading order certain cross sections are written as hard factors multiplying a finite set of process-dependent Wilson-line correlators~\cite{Kotko:2015ura,Marquet:2016cgx}. Process dependence then enlarges the set of operators without, by itself, spoiling predictivity. This holds at leading order for a particular class of processes, and whether it survives at NLO and beyond is, to our knowledge, open. At large $N_c$, the leading-power unpolarized gluon TMDs reduce to dipole and quadrupole correlators~\cite{Dominguez:2011wm,Marquet:2016cgx}. In this limit the evolution closes, with the dipole obeying the BK equation and the quadrupole an equation that involves both the quadrupole and the dipole~\cite{Dominguez:2011gc}. Beyond large $N_c$ the hierarchy no longer closes, and the Gaussian approximation expresses the quadrupole and the other higher-point correlators in terms of the dipole (Sec.~\ref{sec:jimwlk}). 

\subsection{Unpolarised TMD distributions}

TMD distributions naturally enter leading-power factorization theorems for processes characterized by two well-separated scales: a hard scale, which determines the short-distance partonic scattering, and a semi-hard scale, typically of the order of the transverse momentum imbalance, which is sensitive to the intrinsic transverse momentum, $k_\perp$, of partons inside the target. At small $x$, factorization formulae of this type have been derived at leading order, and in some cases at next-to-leading order, for both inclusive and diffractive processes. We begin by reviewing the TMDs relevant for inclusive reactions before turning to diffractive processes. The latter provide particularly sensitive probes of gluon saturation, since the diffractive cross section is quadratic in the scattering amplitude and is therefore dominated by the region where the amplitude is not small, close to the black-disk, or unitarity, limit.

\subsubsection{Inclusive processes}
\label{subsub:inclusive}

We start with the simplest case, which is the TMD distribution of unpolarised gluon inside an unpolarised target. At moderate $x$, the operator definition\footnote{Eq.~\eqref{eq:gTMD-def} is the unsubtracted or ``bare" correlator. A properly defined TMD also requires a rapidity regulator, the corresponding soft-factor subtraction and UV renormalization, which depend on the chosen TMD scheme. They can be ignored at tree level and reappear with the Sudakov logarithms of Sec.~\ref{sec:sudakov}.} is 
\begin{align}
    x G(x,\boldsymbol{k}_\perp)=\int\frac{\der\xi^+\der \boldsymbol{\xi}_\perp}{(2\pi)^3P^-} e^{-ixP^-\xi^+ +i\boldsymbol{k}_\perp\cdot\boldsymbol{\xi}_\perp} 2\langle P|\mathrm{Tr}\big[F^{-i}(\xi^+,\boldsymbol{\xi}_\perp)W(\xi^\mu,0)F^{-i}(0^+,\boldsymbol{0}_\perp)W^\dagger(\xi^\mu,0)\big]|P\rangle\,,\label{eq:gTMD-def}
\end{align}
	where $F^{\mu\nu}=F^{\mu\nu}_at^a$ with $\mathrm{Tr}(t^at^b)=\delta^{ab}/2$, and $W(\xi^\mu,0)$ denotes the gauge link in the fundamental representation connecting the origin to the space-time point $\xi^\mu=(\xi^+,0,\boldsymbol{\xi}_\perp)$ in light-cone coordinates.\footnote{\label{fn:conventions} Throughout this chapter we use light-cone coordinates $x^\pm=(x^0\pm x^3)/\sqrt{2}$, bold symbols for transverse vectors and $\epsilon^{12}=+1$. The target moves along the negative $z$ direction with large momentum $P^-$, so that it couples to the projectile through the field $A^-$ and the Wilson lines are ordered along $x^+$. The projectile moves along the positive $z$ direction, and helicities are defined with respect to the direction of motion of each particle. Both this convention and the opposite one, in which the target moves along the positive $z$ direction, are common in the small-$x$ literature, and the latter is used in particular in much of the work on small-$x$ spin physics~\cite{Kovchegov:2015pbl,Cougoulic:2022gbk,Kovchegov:2025gcg}. The two are related by interchanging the plus and minus components, together with the orientation conventions for the transverse plane and for $\gamma^5$.} The gauge links ensure the gauge invariance of the bilocal operator and resum the soft initial- and final-state interactions of the hard process. Already at moderate $x$, their path was shown to depend on the color flow of the process, from future- or past-pointing staples in SIDIS and Drell--Yan to more complicated paths with closed loops when colored partons are present in both the initial and the final state~\cite{Collins:2002kn,Ji:2002aa,Belitsky:2002sm,Bomhof:2006dp}. This is the origin of the process dependence of TMDs. Taking the small-$x$ limit of Eq.~\eqref{eq:gTMD-def} therefore leads to several distinct gluon TMDs, each characterized by a different gauge-link structure \cite{Dominguez:2010xd,Dominguez:2011wm}. 

\paragraph{Weizs\"acker--Williams and dipole gluon TMDs.}

The two gluon TMDs that play a central role in small-$x$ physics are the WW and dipole gluon distributions~\cite{Kovchegov:1998bi,Dominguez:2011wm}. They are distinguished by the gauge-link structure appearing in their operator definitions.

\begin{figure}[t]
\centering
\includegraphics[width=0.95\textwidth]{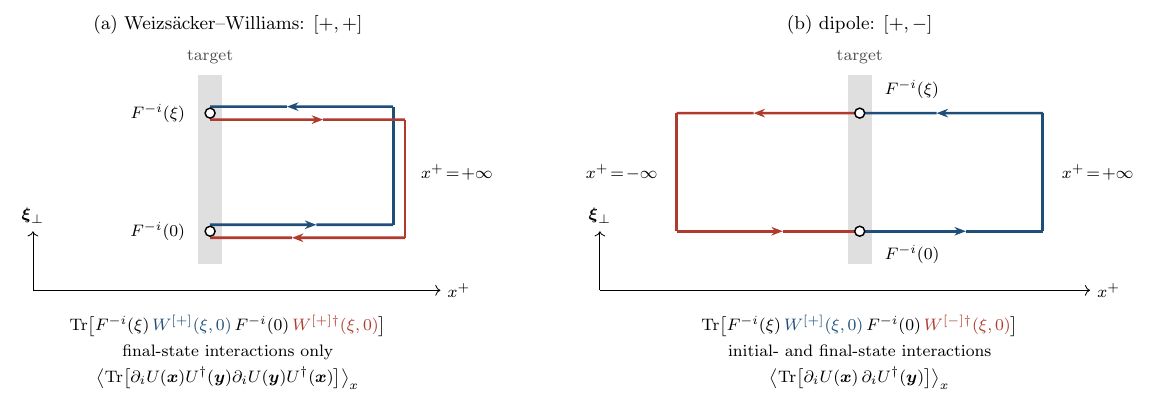}
\caption{Gauge-link structures of the two gluon TMDs that dominate small-$x$ physics, in the $(x^+,\boldsymbol{\xi}_\perp)$ plane, with the target shockwave at $x^+=0$. Each correlator contains two staple-shaped Wilson links in the fundamental representation: $W(\xi,0)$ (blue), running from $0$ to $\xi$, and $W^\dagger(\xi,0)$ (red), running back from $\xi$ to $0$, each through light-cone infinity. (a) The WW distribution: both staples extend to $x^+=+\infty$, so that the active gluon undergoes only final-state interactions. (b) The dipole distribution: one staple extends to $+\infty$ and the other to $-\infty$, so that both initial- and final-state interactions are present. In the small-$x$ limit the links crossing the shockwave become the Wilson lines of Eq.~\eqref{eq:Wilson-line}, and the correlators written below the panels, Eqs.~\eqref{eq:WW-TMD} and~\eqref{eq:dip-CGC}, follow.}
\label{fig:gaugelinks}
\end{figure}

The WW gluon TMD is obtained by choosing two future-pointing staple gauge links, or two past-pointing ones (Fig.~\ref{fig:gaugelinks}a), corresponding to a situation where the active gluon undergoes only final-state (or only initial-state) interactions. The two choices give the same unpolarized distribution by time-reversal invariance but opposite signs for the $T$-odd distributions. In the small-$x$ limit, the field-strength tensors become attached to infinite light-like Wilson lines, allowing the operator to be rewritten entirely in terms of Wilson lines. Using the relation between adjoint and fundamental Wilson lines, the operator takes the form
\begin{equation}
xG_{\rm WW}(x,\boldsymbol{k}_\perp)
=\frac{-2}{\alpha_s}
\int \frac{\der^2\boldsymbol{x}\der^2\boldsymbol{y}}{(2\pi)^4} \ 
e^{-i\boldsymbol{k}_\perp\cdot(\boldsymbol{x}-\boldsymbol{y})}
\left\langle
\mathrm{Tr}
\left[
\partial_i U(\boldsymbol{x})
U^\dagger(\boldsymbol{y})
\partial_i U(\boldsymbol{y})
U^\dagger(\boldsymbol{x})
\right]
\right\rangle_x,\label{eq:WW-TMD}
\end{equation}
where $U(\boldsymbol{x})$ defined as 
\begin{align}
    U(\boldsymbol{x})&=\mathcal{P} \exp\left(ig\int_{-\infty}^{\infty}\der x^+ \ A_a^-(x^+,0,\boldsymbol{x})t^a\right)\,,\label{eq:Wilson-line}
\end{align}
denotes a fundamental Wilson line at transverse position $\boldsymbol{x}$ and $\langle\cdots\rangle_x$ represents the average over the target color fields. The precise meaning of the $x$-dependence of this average will be given in Sec.~\ref{sec:jimwlk}. We work in a gauge in which the target field has only a minus component, such as the covariant gauge, so that the transverse gauge links at light-cone infinity are trivial. In the target light-cone gauge $A^-=0$, these interactions are instead carried by the transverse gauge links~\cite{Ji:2002aa,Belitsky:2002sm}.
 
The WW gluon TMD thus probes the transverse correlations of the gluon field in the target and admits a simple interpretation as the gluon number density in light-cone gauge \cite{McLerran:1993ni,McLerran:1993ka,Kovchegov:1998bi,Iancu:2000hn}. It appears naturally in processes where the hard probe directly resolves the transverse momentum of gluons, most notably in dijet production in deep inelastic scattering (DIS)~\cite{Dominguez:2010xd,Dominguez:2011wm}, which we discuss further in Sec.~\ref{sec:pheno}.

The dipole gluon TMD corresponds instead to a gauge-link structure involving one future-pointing and one past-pointing staple (Fig.~\ref{fig:gaugelinks}b), reflecting the simultaneous presence of initial- and final-state interactions. In the high-energy limit, its operator definition simplifies to a correlator of two fundamental Wilson lines,
\begin{align}
xG_D(x,\boldsymbol{k}_\perp)
&=
\frac{2}{\alpha_s}
\int \frac{\der^2\boldsymbol{x}\,\der^2\boldsymbol{y}}{(2\pi)^4}\,
e^{-i\boldsymbol{k}_\perp\cdot(\boldsymbol{x}-\boldsymbol{y})}
\left\langle
\mathrm{Tr}
\left[
\partial_i U(\boldsymbol{x})\,
\partial_i U^\dagger(\boldsymbol{y})
\right]
\right\rangle_x \nonumber\\
&=\frac{2k_\perp^2}{\alpha_s}
\int \frac{\der^2\boldsymbol{x}\,\der^2\boldsymbol{y}}{(2\pi)^4}\,
e^{-i\boldsymbol{k}_\perp\cdot(\boldsymbol{x}-\boldsymbol{y})}
\left\langle
\mathrm{Tr}
\left[
U(\boldsymbol{x})
U^\dagger(\boldsymbol{y})
\right]
\right\rangle_x\,.
\label{eq:dip-CGC}
\end{align}

The object $S(\boldsymbol{r})=\frac{1}{N_c}\langle\mathrm{Tr}[U(\boldsymbol{0})U^\dagger(\boldsymbol{r})]\rangle_x$ has a simple physical meaning: it is the amplitude for a quark-antiquark pair of transverse size $r_\perp=|\boldsymbol{r}|$ to traverse the target without net color exchange. It tends to one for pairs much smaller than $1/Q_s$, which is color transparency, and to zero for pairs much larger than $1/Q_s$, which is the black-disk limit, and the dipole gluon distribution is essentially its Fourier transform weighted by $k_\perp^2$. The dipole gluon TMD is therefore directly related to the color dipole scattering amplitude $1-S(\boldsymbol{r})$~\cite{Mueller:1989st,Nikolaev:1990ja}, the fundamental building block of high-energy scattering in the CGC framework. It naturally appears in observables involving a color dipole, such as inclusive DIS, forward photon--jet production \cite{Gelis:2002ki,Kolbe:2020tlq}, and forward Drell--Yan production \cite{Gelis:2006hy}.

The gauge-link structures of the two distributions are compared in Fig.~\ref{fig:gaugelinks}. In both cases the operator of Eq.~\eqref{eq:gTMD-def} contains two staple-shaped links in the fundamental representation, $W(\xi,0)$ and $W^\dagger(\xi,0)$. Each runs along $x^+$ to light-cone infinity, crosses the transverse plane there, and returns to the other field strength. The superscript $[\pm]$ labels a staple running to $x^+=\pm\infty$, and the field strengths are localized in the target, which at high energy is a thin shockwave at $x^+=0$. For the WW distribution both staples run to $+\infty$, so that only final-state interactions are resummed, and in the small-$x$ limit the correlator becomes the four-Wilson-line trace of Eq.~\eqref{eq:WW-TMD}. For the dipole distribution one staple runs to $-\infty$, so that both initial- and final-state interactions are resummed, and the correlator reduces to the two-Wilson-line trace of Eq.~\eqref{eq:dip-CGC}.

While the WW and dipole gluon TMDs coincide in the dilute regime, $k_\perp\gg Q_s$, where multiple scattering effects are suppressed, they differ markedly in the saturation region \cite{Dominguez:2011wm,Marquet:2016cgx}, as shown in Table~\ref{tab:TMD-MV-asymptotics}. Their distinction provides one of the clearest manifestations of the process dependence of gluon TMDs: different gauge-link structures give rise to different Wilson-line operators in the CGC, each encoding complementary information about the non-linear gluon dynamics of the target.

\paragraph{Other gluon TMDs.}

In more complicated $pA$ processes involving colored partons in both the initial and final states, the gauge-link structure becomes considerably richer than for the WW and dipole gluon TMDs. Besides future- and past-pointing Wilson lines, the operator definitions contain closed Wilson loops, reflecting the more intricate color flow of the hard scattering \cite{Bomhof:2006dp,Kotko:2015ura}. As discussed at the beginning of this section, such processes violate standard TMD factorization at moderate $x$, whereas at small $x$, in the correlation limit and at leading order, the cross section can be written in terms of a larger set of gluon TMDs, each corresponding to a distinct Wilson-line operator in the CGC~\cite{Dominguez:2011wm,Kotko:2015ura}. The appearance of Wilson loops enriches the color structure of these operators, leading to multi-point correlators beyond the simple dipole and quadrupole operators encountered for the WW and dipole gluon TMDs. A comprehensive discussion of gluon TMDs beyond the WW and dipole distributions can be found in~\cite{Petreska:2018cbf}, including their operator definitions, their expressions in the CGC framework in terms of light-like Wilson lines, and their analytic forms in the Golec-Biernat--W\"{u}sthoff~\cite{Golec-Biernat:1998zce} and McLerran--Venugopalan models~\cite{McLerran:1993ni,McLerran:1993ka} for the dipole scattering amplitude.

\paragraph{(Sea) quark TMDs.} 

At small $x$, the dominant quark contribution arises from sea quarks generated through gluon splitting. As in the gluon sector, the gauge-link structure reflects the color flow of the hard process. The simplest quark TMD is the one appearing in SIDIS and, by time-reversal symmetry, in the Drell--Yan process, where the operator definitions differ only by the direction of the staple gauge links. In the CGC formalism, both operators can be expressed in terms of Wilson lines and are closely related to the scattering of a quark-antiquark dipole off the target~\cite{Marquet:2009ca,Xiao:2010sa}. Beyond these benchmark processes, more complicated reactions such as forward dijet production in $pA$ collisions involve a richer zoology of sea-quark TMDs with additional gauge-link structures generated by multiple initial- and final-state interactions. Remarkably, this proliferation of distributions has a simple underlying structure: at large $N_c$, all leading-power sea-quark TMDs can be constructed from two building blocks, namely the sea-quark TMD probed in SIDIS and Drell--Yan and the dipole $S$-matrix~\cite{Caucal:2025xxh}. The process dependence is therefore entirely encoded in the way these two operators combine, in close analogy with the gluon sector. The additional rescattering effects generally probe a larger effective saturation scale in hadronic processes than in SIDIS (see Fig.~\ref{fig:seaquark}). This emerging generalized universality to two building blocks, which holds at leading order and at large $N_c$, considerably simplifies the description of sea-quark TMDs at small $x$.

\subsubsection{Diffractive processes}

Diffractive processes in which a rapidity gap is required in the final state are of particular importance at small $x$, as they are especially sensitive to gluon saturation. The existence of a rapidity gap can be physically interpreted as resulting from the exchange of a colourless object between the dilute probe and the dense target, known as a Pomeron. Whereas diffractive parton distributions are extracted from data through fits within collinear factorization, at small $x$ the CGC framework allows one to compute the longitudinal and transverse momentum distributions of partons inside the Pomeron directly from the dipole amplitude, with saturation built in.

These features make diffractive processes a particularly useful setting for studying the interplay between TMD factorization and saturation effects. In this context, significant progress has recently been made in establishing TMD factorization for diffractive processes in the small-$x$ regime. These include semi-inclusive DIS~\cite{Hatta:2022lzj}, diffractive dijet production in DIS~\cite{Hauksson:2024bvv}, and diffractive dijet production in ultra-peripheral collisions (UPCs)~\cite{Iancu:2023lel}. As in the inclusive case, the relevant quark and gluon TMDs admit two complementary derivations. On the one hand, one can start from their operator definitions and take the high-energy limit to obtain their representation in terms of Wilson-line correlators within the CGC~\cite{Hatta:2022lzj,Bhattacharya:2025fnz,Bhattacharya:2026bal}. On the other hand, one may compute the corresponding diffractive process at leading order in the CGC framework and subsequently perform a correlation-limit expansion, in which the hard scale is separated from the semi-hard transverse momentum imbalance. Matching the resulting expression onto a TMD factorization formula then identifies the diffractive TMDs and their associated Wilson-line operators. 

The benefit of the observable-based approach is that it provides a more transparent physical interpretation of the factorization formula, and, in particular, clarifies how a given diffractive TMD enters at leading power in a given process. For example, consider diffractive dijet production in DIS. At leading order, the cross section would, in principle, be given by the square of the $\gamma^* \to q\bar q$ amplitude, with the quark-antiquark pair produced in a colour-singlet state. It turns out that, in the back-to-back limit, where the relative transverse momentum of the pair $P_\perp$ is much larger than its transverse momentum imbalance $k_\perp$, the cross section falls as $1/P_\perp^6$ and is therefore power-suppressed at large $P_\perp$, compared, for example, with the inclusive dijet cross section at LO, which falls as $1/P_\perp^4$.

The reason is that, when the $q\bar q$ pair is hard, it corresponds, by the Heisenberg principle, to a very small colour-singlet dipole in coordinate space. This colour-singlet dipole therefore becomes transparent and interacts only very weakly with the target. In order to obtain leading-power TMD factorization, one must therefore go to next-to-leading order~\cite{Boussarie:2016ogo,Fucilla:2023mkl,Fucilla:2022wcg} and consider an additional gluon emission. When this gluon is relatively soft, with a longitudinal momentum fraction of the projectile virtual-photon momentum of the order of the ratio between the imbalance and the hard scale squared~$\sim k_\perp^2/P_\perp^2$, and with transverse momentum of the order of the $q\bar q$ pair's imbalance $k_\perp$, a large effective gluon-gluon dipole is formed, with the other effective gluon being built from the small $q\bar q$ pair. This large gluon-gluon dipole, in a colour-singlet state, can now interact strongly with the target and gives a sizeable, leading-power contribution to the cross section~\cite{Iancu:2021rup,Iancu:2022lcw,Hauksson:2024bvv}, which takes the form
\begin{align}
    \frac{\der\sigma_{\rm diff.}^{\gamma^*\to q\bar q (g)}}{\der^2\Pt\der^2\kt\der y_q\der y_{\bar q}\der\ln(1/x_{\mathbb{P}})}&=\alpha_s H_{\rm LO}(\Pt,Q,y_q,y_{\bar q}) \times x G_{\mathbb{P}}(x,x_{\mathbb{P}},\kt) \,,
\end{align}
where $H_{\rm LO}(\Pt,Q,y_q,y_{\bar q})$ is the same leading-order hard factor as in the inclusive case, which depends on the hard scales $P_\perp$ and $Q$ and on the rapidities of the produced jets, and the gluon diffractive TMD reads in the CGC
\begin{align}
          x G_{\mathbb{P}}(x,x_{\mathbb{P}},\kt)&=\frac{(N_c^2-1) S_\perp}{8\pi^4(1-x)}\left[\Omega^2\int_0^\infty \der r\, r\, \operatorname{J}_2(k_\perp r)\operatorname{K}_2(\Omega r)\left(1-\frac{1}{N_c^2-1}\langle\textrm{Tr}\,\mathcal{U}(\boldsymbol{0}_\perp)\mathcal{U}^\dagger(\boldsymbol{r})\rangle_{x_{\mathbb{P}}}\right)\right]^2 \,, \nonumber\\
\Omega^2&=\frac{x}{1-x}k_\perp^2 \,, \label{eq:gluon-diffTMD}
\end{align}
where $x$ is the longitudinal momentum fraction of the exchanged gluon relative to the Pomeron, $x_{\mathbb{P}}$ is the fraction of the target momentum carried by the Pomeron, and $S_\perp$ is the transverse area of the target. Here $J_n$ and $K_n$ are respectively Bessel functions of the first kind and modified Bessel functions of the second kind. More refined expressions including the dependence on the total momentum transfer from the target can be found in~\cite{Rodriguez-Aguilar:2024efj} for incoherent diffraction and also in~\cite{Caucal:2026xhj} in momentum space. As a result of the effective gluon-gluon dipole appearing in the amplitude, this expression features the small-$x$ dipole correlator, but in the adjoint representation with $\mathcal{U}$ adjoint Wilson lines. Eq.\,\eqref{eq:gluon-diffTMD} provides a noteworthy example of an explicit expression for the $k_\perp$ and $x$ dependence of the gluon inside the Pomeron, which itself carries a longitudinal momentum fraction $x_{\mathbb{P}}$ of the target. Using the McLerran–Venugopalan model for the dipole correlator, one can show that, at leading order, $x G_{\mathbb{P}}(x,x_{\mathbb{P}},\kt)$ falls as $1/k_\perp^4$ at large $k_\perp$, rather than exhibiting the typical $1/k_\perp^2$ tail of gluon TMDs in inclusive processes (see Table~\ref{tab:TMD-MV-asymptotics}). Consequently, the dominant contribution to the cross section indeed comes from the saturated regime, $k_\perp \lesssim Q_s$.

Similar results have been recently obtained for the quark diffractive TMDs at small $x$, which enter diffractive single-inclusive hadron production in DIS in the limit $Q^2\gg k_\perp^2$ or diffractive quark+gluon dijet production in DIS in the back-to-back limit, for massless and massive quark cases~\cite{Hatta:2022lzj,Hauksson:2024bvv,Caucal:2026xhj}.

\begin{table}[t]
    \centering
    \small
    \renewcommand{\arraystretch}{1.4}
    \setlength{\tabcolsep}{7pt}

    \begin{tabular}{
        >{\centering\arraybackslash}m{2.3cm}
        >{\centering\arraybackslash}m{2.5cm}
        >{\centering\arraybackslash}m{2.5cm}
        >{\centering\arraybackslash}m{4.5cm}
    }
        \rowcolor{mvblue}
        \multicolumn{4}{c}{
            \color{white}\normalsize\bfseries
            Small- and large-$k_\perp$ behavior of TMDs in the MV model
        } \\

        \rowcolor{mvbluelight}
        \textbf{TMD}
        & \textbf{$k_\perp\ll Q_s$}
        & \textbf{$k_\perp\gg Q_s$}
        & \textbf{Process} \\

        \midrule

        \rowcolor{mvbluerow}
        \textbf{WW gluon TMD}
        &
        $\displaystyle \ln\frac{Q_s^2}{k_\perp^2}$
        &
        $\displaystyle \frac{Q_s^2}{k_\perp^2}$
        &
        Dijet production in DIS, Higgs and quarkonium-pair production in $pA$
        \\

        \addlinespace[2pt]

        \textbf{Dipole gluon TMD}
        &
        $\displaystyle \frac{k_\perp^2}{Q_s^2}$
        &
        $\displaystyle \frac{Q_s^2}{k_\perp^2}$
        &
        Inclusive DIS, forward photon--jet and Drell--Yan production in $pA$
        \\

        \addlinespace[2pt]

        \rowcolor{mvbluerow}
        \textbf{Sea-quark TMD}
        &
        $\mathrm{const.}$
        &
        $\displaystyle \frac{Q_s^2}{k_\perp^2}\ln\frac{k_\perp^2}{\Lambda^2}$
        &
        SIDIS, Drell--Yan, lepton--jet correlations in DIS, forward dijet production in $pA$
        \\

        \addlinespace[2pt]

        \textbf{Gluon diffractive TMD}
        &
        $\mathrm{const.}$
        &
        $\displaystyle \frac{\tilde Q_{s,g}^4}{k_\perp^4}\ln^2\frac{k_\perp^2}{\Lambda^2}$
        &
        Diffractive $2+1$ jet production in DIS and in UPCs
        \\

        \addlinespace[2pt]

        \rowcolor{mvbluerow}
        \textbf{Quark diffractive TMD}
        &
        $\mathrm{const.}$
        &
        $\displaystyle \frac{\tilde Q_{s}^4}{k_\perp^4}\ln^2\frac{k_\perp^2}{\Lambda^2}$
        &
        Diffractive SIDIS, diffractive dijets with a soft quark
        \\

        \bottomrule
    \end{tabular}

    \caption{Small- and large-$k_\perp$ behavior of the gluon, sea-quark and diffractive TMDs at tree level in the McLerran--Venugopalan (MV) model, up to logarithms absorbed in the saturation scale, with $\Lambda$ the infrared cutoff of the model~\cite{Dominguez:2011wm,Marquet:2009ca,Iancu:2021rup,Hauksson:2024bvv}. For the diffractive distributions, $x$ is the momentum fraction relative to the Pomeron and the effective saturation scales are $\tilde Q_s^2=(1-x)Q_s^2$ and $\tilde Q_{s,g}^2=(1-x)Q_{s,g}^2$, with $Q_{s,g}^2\simeq(N_c/C_F)Q_s^2$ the saturation scale probed by a gluon dipole.}
    \label{tab:TMD-MV-asymptotics}
\end{table}

\subsubsection{Target fragmentation region and fracture functions}

Another interesting kinematic regime, intermediate between the inclusive and diffractive cases and specific to the DIS process is the so-called target fragmentation region (TFR), which has received attention recently, and could be of interest for the EIC. When one measures a hadron or a jet in the target hemisphere of the Breit frame, rather than in the current one, the appropriate description is not in terms of parton distributions and fragmentation functions separately. One works instead with (extended) \emph{fracture functions}, which describe the joint probability of resolving a parton of momentum fraction $x$ and of producing a hadron or jet with longitudinal momentum fraction $\xi=k^-/P^-$ of the target and transverse momentum $k_\perp$. These objects were introduced and developed within collinear factorization~\cite{Trentadue:1993ka,Grazzini:1997ih,Berera:1995fj}. Diffractive parton distributions~\cite{Collins:1997sr,Berera:1994xh,Hautmann:1999ui} are the special case in which the observed ``hadron'' is the intact target and in which a rapidity gap is required in the final state. Although the ``spectator'' parton does not participate directly in the hard scattering, it retains information about the internal partonic structure of the parent hadron and, in particular, can be sensitive to saturation in nuclei.

At small $x$, a particle measured in the target hemisphere is still separated from the target by a large rapidity interval. It is a spectator parton produced in the perturbative splitting of a target gluon, and the target fragmentation region can be described in terms of lightlike Wilson-line correlators. One may therefore ask whether a TMD factorization theorem holds there as well. What has been derived is a factorization in terms of extended fracture functions, which are expressed in terms of CGC operators~\cite{Caucal:2025qjg}. For single-inclusive jet production by a longitudinally polarized photon with $k_\perp\ll Q$, the leading-power cross section arises at next-to-leading order in $\alpha_s$ and factorizes onto extended quark and gluon jet fracture functions. At this order the quark one coincides with the sea-quark TMD of Sec.~\ref{subsub:inclusive}, and the gluon one involves the adjoint dipole. The quark fracture function saturates below $Q_s$ at a value independent of the nuclear mass number, whereas the gluon one keeps its $1/k_\perp^2$ tail down to small $k_\perp$ and is nearly insensitive to saturation~\cite{Caucal:2025qjg}. This is of interest for the EIC, although the phase space for jets with $k_\perp\ll Q$ at small $x$ is narrow there. It also closes a gap in the classification, since the inclusive, diffractive and target-fragmentation observables are then all described by objects built from the same small set of Wilson-line operators, differing only in which of them the final state selects. We note that this analysis was carried out at leading order in the small-$x$ limit and for $\xi$-integrated functions, and that the corresponding evolution has not yet been worked out. The extension to identified hadrons is discussed in Sec.~\ref{sec:recoil}~\cite{Caucal:2026zpt}.

Furthermore, fracture functions are related to the nucleon energy correlators~\cite{Liu:2022wop,Liu:2023aqb}, which measure the angular distribution of energy flow in the target: a sum rule expresses the energy-pattern structure functions of this region, up to twist three, through the associated correlators~\cite{Chen:2024bpj}. At small $x$ the gluonic correlators have been computed at eikonal accuracy in the shockwave background~\cite{Mantysaari:2026zte}, where the linearly polarized gluon correlator generates a $\cos2\phi$ asymmetry in the DIS energy pattern, governed by its ratio to the unpolarized quark correlator, which is sensitive to saturation and shows a substantial nuclear suppression.

\subsection{Polarised TMD distributions}

We now turn to the polarized sector and consider two interesting cases. The first concerns the polarization of the parton itself inside an unpolarized target, of which the most important example at small $x$ is the distribution of linearly polarized gluons. The second concerns the polarization of the target, which selects TMDs such as the Sivers function. As we shall see, the two cases behave rather differently in the high-energy limit: the linearly polarized gluon distributions survive at leading eikonal accuracy and are as large as their unpolarized counterparts, whereas most target-polarization-dependent distributions are either eikonally suppressed or reduce, at leading power, to a single $C$-odd operator. Distributions that appear only beyond eikonal accuracy, and in particular the helicity distributions, are deferred to Sec.~\ref{sec:beyond-eikonal}.

\subsubsection{Linearly polarized gluons in an unpolarized target}

For an unpolarized target, the leading-twist gluon correlator $\Gamma^{ij}(x,\boldsymbol{k}_\perp)$ (defined as the generalization of Eq.\,\eqref{eq:gTMD-def} with open $i,j$ indices) admits the decomposition
\begin{equation}
\Gamma^{ij}(x,\boldsymbol{k}_\perp)
=\frac{1}{2}\,\delta^{ij}\, xG(x,\boldsymbol{k}_\perp)
+\frac{1}{2}\left(\frac{2 k^i_\perp k^j_\perp}{\boldsymbol{k}_\perp^2}-\delta^{ij}\right) x h_1^{\perp\,g}(x,\boldsymbol{k}_\perp)\,,
\label{eq:linpol-decomposition}
\end{equation}
where $xG$ is the unpolarized gluon TMD discussed above and $x h_1^{\perp\,g}$ is the distribution of linearly polarized gluons~\cite{Metz:2011wb,Dominguez:2011br}. Both terms are leading twist and, at small $x$, both are expected to survive the eikonal $x\to 0$ limit. Each carries its own gauge-link structure, so that Eq.~\eqref{eq:linpol-decomposition} applies separately to the WW and to the dipole correlators.

The two cases are however qualitatively different. For the dipole correlator the two derivatives in the first form of Eq.~\eqref{eq:dip-CGC} act on different Wilson lines, so that after integration by parts they act only on the Fourier phase, and the correlator with open indices $i,j$ reduces to $k_\perp^i k_\perp^j$ times the Fourier transform of the dipole $S$-matrix. Since $k_\perp^ik_\perp^j$ has equal trace and traceless parts in the decomposition of Eq.~\eqref{eq:linpol-decomposition}, the two dipole distributions coincide. The equality
\begin{equation}
x h_{1,D}^{\perp\,g}(x,\boldsymbol{k}_\perp)=xG_D(x,\boldsymbol{k}_\perp) \,,
\label{eq:h1dip}
\end{equation}
therefore holds exactly for the tree-level correlator, for any target and at all transverse momenta~\cite{Metz:2011wb,Dominguez:2011br}, and it is preserved by JIMWLK evolution, which acts on the dipole $S$-matrix alone: the gluons probed by a dipole are maximally linearly polarized. The equality holds for the eikonal operator and its evolution in $x$. It is broken by sub-eikonal corrections to the operator, and by the evolution in the hard scale: at one loop in $\alpha_s$, the real emission generates the unpolarized distribution from the collinear gluon density with a logarithm of the hard scale, but the linearly polarized one without it, so that the degree of polarization decreases toward higher scales~\cite{Sun:2011iw,Boer:2014tka,Echevarria:2015uaa}.  

For the WW correlator the situation is richer. While $x h_{1,\rm WW}^{\perp\,g}$ approaches $xG_{\rm WW}$ at large $k_\perp\gg Q_s$, where the target is dilute, it is strongly suppressed relative to $xG_{\rm WW}$ in the saturation region $k_\perp\lesssim Q_s$~\cite{Metz:2011wb,Dumitru:2015gaa}. The ratio $x h_{1,\rm WW}^{\perp\,g}/xG_{\rm WW}$ could therefore serve as a measure of the onset of saturation, and it is major theoretical motivation for measuring azimuthal asymmetries in back-to-back dijet production.

Phenomenologically, the linearly polarized distributions generate a $\cos(2\phi)$ modulation of the azimuthal angle between the dijet total and relative transverse momenta. This modulation has been studied for dijets in DIS~\cite{Metz:2011wb,Dumitru:2015gaa,Dumitru:2018kuw,Mantysaari:2019hkq}, and for heavy-quark pair production in $pA$ collisions~\cite{Marquet:2017xwy}. An important caveat, to which we return in Sec.~\ref{sec:pheno}, is that soft-gluon radiation also generates azimuthal anisotropies of the same functional form~\cite{Hatta:2021jcd}, so that any quantitative extraction of $x h_1^{\perp\,g}$ requires the Sudakov contribution to be under control. Linearly polarized gluons have also been found to enhance the entanglement of back-to-back heavy-quark pairs produced in electron--nucleus collisions when the total and relative transverse momenta of the pair are orthogonal~\cite{Fucilla:2026mkg}, connecting small-$x$ TMDs with quantum-information observables.

\subsubsection{Transversely polarized targets: the Sivers function and the odderon}
\label{subsub:sivers}

For a transversely polarized target, the correlator acquires an additional structure proportional to $\epsilon^{ij}_\perp k^i_\perp S^j_\perp$, which defines the gluon Sivers function $f_{1T}^{\perp\,g}$. Since the Sivers function is $T$-odd, it requires both a non-trivial gauge link and an imaginary part of the Wilson-line correlator. The WW correlator is real, so that the WW-type Sivers function receives no eikonal contribution and is suppressed by powers of $x$, and at leading power only the dipole-type function survives.

The dipole-type function is the expectation value of a single Wilson loop, the spin-dependent odderon, the $C$-odd imaginary part of the dipole correlator of Eq.~\eqref{eq:dip-CGC},
\begin{equation}
O(\boldsymbol{b},\boldsymbol{r})
=\frac{1}{2iN_c}\left\langle
\mathrm{Tr}\left[U(\boldsymbol{x})U^\dagger(\boldsymbol{y})\right]
-\mathrm{Tr}\left[U(\boldsymbol{y})U^\dagger(\boldsymbol{x})\right]
\right\rangle\,,
\qquad
\boldsymbol{b}=\frac{\boldsymbol{x}+\boldsymbol{y}}{2}\,,\quad \boldsymbol{r}=\boldsymbol{x}-\boldsymbol{y}\,,
\label{eq:odderon}
\end{equation}
whose non-vanishing value requires a correlation between $\boldsymbol{b}$ and $\boldsymbol{S}_\perp$. Furthermore, the same operator exhausts the $T$-odd gluon sector of a transversely polarized target: $h_1^{g}$ and $h_{1T}^{\perp\,g}$, which at moderate $x$ are independent of the Sivers function, reduce to it as well at small $x$~\cite{Boer:2015pni,Boer:2016xqr}. The odderon is a long-standing object of study in high-energy QCD in its own right~\cite{Kwiecinski:1980wb,Hatta:2005as,Kovchegov:2003dm,Jeon:2005cf}, and its small-$x$ evolution is governed by the $C$-odd sector of the Balitsky--Kovchegov and JIMWLK equations~\cite{Hatta:2005as,Lappi:2016gqe}. This identification therefore provides a direct connection between a spin observable and a well-defined saturation-physics operator. In a first numerical study of the small-$x$ gluon Sivers function, the odderon evolution equation was solved in both its linear and non-linear regimes~\cite{Yao:2018vcg}. It was found that the $x$ and $k_\perp$ dependences do not factorize once quantum evolution is included, although the resulting factorization breaking appears to be numerically mild. 

The non-perturbative initial condition for this evolution is less constrained. In the standard MV model the color charges are Gaussian, and the expectation value of the $C$-odd operator vanishes. A non-zero odderon requires a cubic term $d^{abc}\rho^a\rho^b\rho^c$ in the MV weight functional~\cite{Jeon:2005cf}, and for a transversely polarized proton it is generated by an azimuthally asymmetric transverse distribution of the valence quarks~\cite{Zhou:2013gsa}. In a quark-diquark model, the same asymmetry of the color sources ties the spin-dependent odderon to parton orbital angular momentum~\cite{Szymanowski:2016mbq}. The dipole gluon Sivers function has since been evaluated at moderately small $x$ in a three-quark light-front model, together with the BFKL anomalous dimension that controls its large-$k_\perp$ tail~\cite{Benic:2026nnf}. While this model is formulated at moderate $x$, it provides a first estimate of the normalization, which could serve as an initial condition for the evolution.

The same operator controls the flavor non-singlet quark Sivers function, whose leading, eikonal contribution at small $x$ is again the spin-dependent odderon~\cite{Kovchegov:2021iyc}. At leading power in $x$, the non-vanishing $T$-odd distributions, namely the quark Sivers function and the dipole-type gluon ones, are thus controlled by one $C$-odd Wilson-loop operator, with all process dependence residing in how that operator is convoluted with the hard factors.

Evidence for $C$-odd exchange in elastic $pp$ and $p\bar p$ scattering has been reported by the D0 and TOTEM collaborations~\cite{D0:2020tig}, but the spin-dependent odderon that enters here has not been measured. Concrete proposals exist:  exclusive pion production in unpolarized electron-proton scattering in the forward
region~\cite{Boussarie:2019vmk}, transverse energy–energy correlators for nearly back-to-back hadron–electron pairs off a transversely polarized proton at the EIC, whose asymmetry is directly sensitive to the odderon amplitude because the spin dependence enters through a $C$-odd exchange, with an estimated size of $10^{-4}$ \cite{Bhattacharya:2025bqa}, and track-based nucleon energy correlators in the target fragmentation region, whose single-spin asymmetry changes sign between positively and negatively charged hadrons \cite{Mantysaari:2025mht}.

We return to the polarized distributions, which form a rich subject beyond eikonal accuracy, in Sec.~\ref{sec:beyond-eikonal}.

\section{Evolution equations for small $x$ gluon and quark TMDs}

\label{sec:evolution}

In the previous section, we have explored the variety of QCD operators describing TMD quark and gluon distributions in protons and nuclei and how the small $x$/CGC formalism enables one to model their transverse momentum dependence at leading order from perturbative QCD thanks to the emergent semi-hard saturation scale $Q_s(x)\gtrsim \Lambda_{\rm QCD}$. This section is devoted instead to quantum corrections, and in particular renormalization-group (RG) equations satisfied by these operators in the small-$x$ regime. We only consider the TMD distributions which are not power suppressed at small $x$. The case of the subeikonal TMDs will be discussed in Sec.~\ref{sec:beyond-eikonal}, and in particular in subsection~\ref{sub:smallx-helicity}.

\subsection{High-energy evolution: JIMWLK renormalization-group equation}

\label{sec:jimwlk}

Compared to TMD factorization at moderate $x$, a key challenge in the small-$x$ regime is that the perturbative expansion of TMD parton distributions is plagued, to all orders in perturbation theory, by large logarithmic corrections of the form $(\alpha_s\ln(1/x))^n$, $\alpha_s(\alpha_s\ln(1/x))^n$, $\alpha_s^2(\alpha_s\ln(1/x))^n$, and so on. When a given process in which the TMD enters probes values of $x\ll 1$, the smallness of the strong coupling is parametrically compensated by this large $\ln(1/x)$ logarithm, such that a naive truncation of the perturbative expansion eventually breaks down. An all-order resummation of these logarithmically enhanced contributions is therefore essential for obtaining reliable predictions. The resummation of powers of $(\alpha_s\ln(1/x))^n$ defines leading-logarithmic (LL) accuracy, that of $\alpha_s(\alpha_s\ln(1/x))^n$ next-to-leading-logarithmic (NLL) accuracy, and so forth.

The CGC formalism (or, equivalently, the high-energy OPE) applied to dilute-dense collisions provides a consistent theoretical framework for carrying out this resummation program in the non-linear QCD regime, while systematically isolating purely perturbative corrections $\mathcal{O}(\alpha_s^n)$ that are not enhanced by $\ln(1/x)$. More specifically, when a given cross-section that factorizes at leading power in terms of a small-$x$ TMD distribution is computed beyond leading order, the NLO correction contains a logarithmically divergent contribution associated with the region in which the longitudinal momentum of a gluon along the projectile direction becomes small. Schematically, at NLO,
\begin{align}
    \der\sigma &= H_{\rm LO}(Q)\times  \left[F(\kt)+\alpha_s\int\frac{\der k_g^+}{k_g^+}\mathcal{H}_{\rm JIMWLK}\otimes F(\kt)+\mathcal{O}(\alpha_s)\right]+\mathcal{O}(k_\perp/Q,Q_s/Q)\,, \label{eq:slow-div}
\end{align}
where here, $Q$ and $\kt$ generically denote the hard and semi-hard scale such that $Q\gg k_\perp,Q_s$ (with no hierarchy between $k_\perp$ and $Q_s$). $H_{\rm LO}(Q)$ is the LO hard factor encoding the dependence on the hard scale, while $F(\kt)$ is the leading order (bare) TMD associated with a given process, which is expressed within the CGC as an average of an operator built from Wilson lines as discussed in the previous section. The second term in the square bracket is the NLO correction to the cross-section which is logarithmically divergent in the infrared, when $k_g^+\to 0$ (with the plus component being the large momentum component of the dilute projectile). In the UV, $k_g^+$ is always constrained by the kinematic of the probe because of four-momentum conservation. The kernel of this logarithmically divergent contribution is universal, process-independent: it can be shown to correspond to the action of the LL JIMWLK Hamiltonian $\mathcal{H}_{\rm JIMWLK}$~\cite{Iancu:2001ad, Jalilian-Marian:1997qno,Jalilian-Marian:1997ubg,Kovner:2000pt, Iancu:2001md,Ferreiro:2001qy} on the small-$x$ operator $F(\kt)$ entering the leading-order cross section.

This observation allows one to formulate the resummation in terms of a renormalization-group evolution, whereby a factorization scale is introduced to isolate the logarithmically divergent contribution and absorb it into the renormalization of the bare small-$x$ TMD operator.
As we shall see in the next subsection, the gluon phase space that is effectively absorbed into the renormalized TMD and subtracted from the NLO correction plays an important role in the perturbative stability of the cross section and in properly isolating other potentially large logarithms of the ratio $Q/k_\perp$. At this stage of the discussion (and formally, to leading logarithmic accuracy), however, it is sufficient to introduce a factorization scale $k_f^-$ on the minus component of the gluon momentum, parametrically related to $k_g^+$ by the on-shell condition $k_g^-=k_{g\perp}^2/(2k_g^+)$, rather than on $k_g^+$ itself, such that the bare TMD gets replaced by the renormalized one according to
\begin{align}
    F(\kt)\to F(k_f^-,\kt)&\equiv F(\kt)+\alpha_s\int_{\Lambda^-}^{k_f^-}\frac{\der k_g^-}{k_g^-}\mathcal{H}_{\rm JIMWLK}\otimes F(\kt)\nonumber\\
&=F(\kt)+\alpha_s\ln(k_f^-/\Lambda^-)\mathcal{H}_{\rm JIMWLK}\otimes F(\kt) \,,
\end{align}
where $\Lambda^-$ is an arbitrary cut-off. The leading-logarithmic JIMWLK equation for the small-$x$ TMD~$F(k_f^-,\kt)$,
\begin{align}
    k_f^-\frac{\partial F(k_f^-,\kt)}{\partial k_f^-}&=\alpha_s\mathcal{H}_{\rm JIMWLK}\otimes F(k_f^-,\kt) \,, \label{eq:JIMWLK-eq}
\end{align}
then follows from the $k_f^-$ independence of the cross-section. Through these manipulations, the renormalized operator acquires an effective scale dependence through evolution with the $k_f^-$ factorization scale, while the remaining finite, but factorization-scale-dependent, contribution is absorbed into the NLO coefficient function, or impact factor. To minimize the latter, $k_f^-$ must be typically chosen such that $k_f^- = x_g P^-$, where $P^-$ is the large minus component of the nucleon four-momentum and $x_g$ is the gluon longitudinal momentum fraction probed by the process. In the CGC framework, this scale dependence can also be interpreted as a scale dependence of the gauge-invariant weight functional describing the probability distribution of the large-$x$ color charges that source the target background field. This is why the $x$ dependence of small-$x$ TMDs is often made explicit in the notation for the CGC average, $\langle\ldots\rangle_x$ as in Eq.\,\eqref{eq:WW-TMD}.

At a given perturbative order, the physical cross section is independent of the arbitrary factorization scale, since its dependence cancels between the small-$x$ operator and the coefficient function. This framework has been established at leading logarithmic accuracy for a broad class of TMD-factorizable processes in recent years, both in the dilute regime and, more importantly, in the dense regime. An important challenge for the future is to establish high-energy factorization at next-to-leading logarithmic accuracy. This requires either the computation of a small-$x$ TMD in the dense regime at two loops starting from its operator definition, or, equivalently, the computation at two loops of an observable that is expected to factorize at leading power in terms of the corresponding TMD. We also note that, in addition to the observable-based approach followed here to derive the JIMWLK equation for TMDs, the same equation can be obtained at the operator level, without referring to a specific cross section. This was done in~\cite{Balitsky:2015qba,Balitsky:2016dgz} using rapidity-only factorization, and more recently in~\cite{Mukherjee:2023snp}, although only in the dilute regime. A key difficulty with the operator-level approach is that, in the non-linear saturation regime, the JIMWLK equation effectively involves an infinite tower of operators, corresponding to the Balitsky hierarchy. A complete treatment of saturation power corrections with typical parametric size $Q_s/k_\perp$ must therefore keep track of all terms in this hierarchy. For the specific case of the WW gluon TMD, the JIMWLK equation applied to the corresponding CGC operator Eq.\,\eqref{eq:WW-TMD} has been obtained in coordinate space in~\cite{Dominguez:2011gc} and in momentum space in~\cite{Caucal:2025mth}, and it couples the WW gluon TMD with the dipole gluon TMD as well as three-point CGC correlators.

Let us now discuss the effect of JIMWLK evolution on small-$x$ TMDs. The JIMWLK equation is notoriously difficult to solve, although numerical solutions have been obtained for the WW gluon TMD and its linearly polarized partner~\cite{Dumitru:2015gaa}, for all gluon TMDs relevant for DIS and $pA$ processes~\cite{Marquet:2016cgx}, and, more recently, for the quark--gluon and gluon--gluon distributions of forward dijet production~\cite{Cougoulic:2026sww}. In practice, one therefore often relies on the Gaussian approximation, a mean-field truncation of the Balitsky hierarchy in which all higher-point correlators, and hence all TMDs, are expressed through the dipole, whose $x$ dependence is governed by the BK equation. This truncation is controlled in the dilute regime, where it reproduces BFKL dynamics, and deep in the saturation regime~\cite{Iancu:2011ns,Iancu:2011nj}, and numerical solutions of the JIMWLK equation have shown that it describes the quadrupole and other higher-point correlators to good accuracy~\cite{Dumitru:2011vk,Alvioli:2012ba,Lappi:2015vta}. For TMDs the picture is more nuanced. At the MV initial condition, where the color charges are Gaussian, the Gaussian expressions reproduce the full numerical results for the three quark--gluon and seven gluon--gluon distributions of forward dijet production, for $N_c$ between $2$ and $5$, and their large-$N_c$ limit coincides with the mean-field results~\cite{Cougoulic:2026drr}. Once JIMWLK evolution is included, however, the Gaussian truncation reproduces the color structures related to the fundamental and adjoint dipoles but underestimates the speed of evolution of the others, increasingly so for larger color representations. Since the TMDs are linear combinations of these structures, the discrepancies can be amplified in the distributions themselves, at least in position space~\cite{Cougoulic:2026sww}, and their impact on momentum-space distributions and cross sections remains to be quantified. For TMD factorization in incoherent diffractive processes in photon-nucleus collision, the use of the full JIMWLK equation instead of the Gaussian approximation is even more critical as shown in~\cite{Lappi:2026kxi}. The Gaussian approximation should therefore be regarded as a reasonable starting point rather than a controlled approximation for the full set of TMDs, and phenomenological studies that rely on it carry a theoretical uncertainty that has not yet been quantified.

\begin{figure}[t]
\centering
\includegraphics[width=\textwidth]{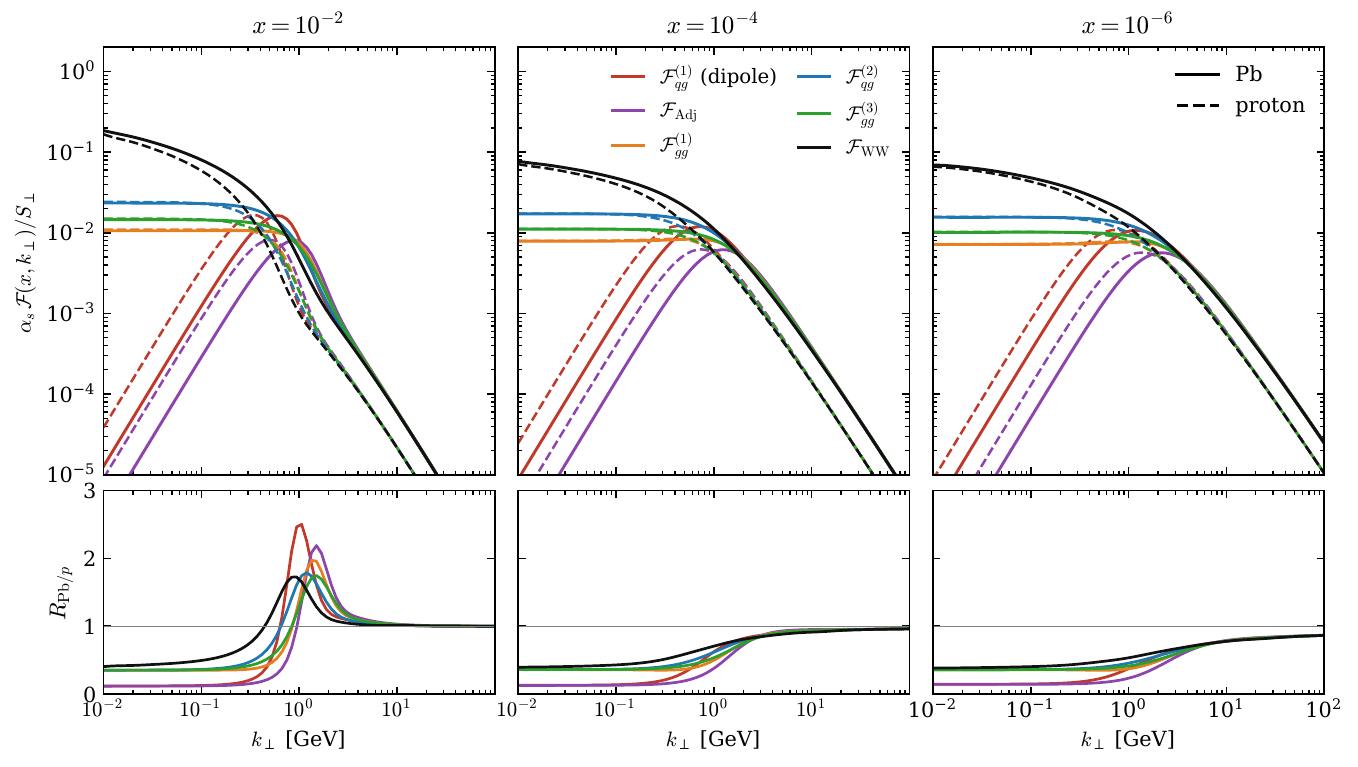}
\caption{Small-$x$ gluon TMDs, normalized by the transverse area $S_\perp$, for a lead nucleus (solid) and a proton (dashed) at $x=10^{-2}$, $10^{-4}$ and $10^{-6}$. $\mathcal{F}^{(1)}_{qg}$ and $\mathcal{F}_{\rm WW}$ are the dipole and WW distributions of Eqs.~\eqref{eq:dip-CGC} and~\eqref{eq:WW-TMD}, $\mathcal{F}_{\rm Adj}$ is the adjoint dipole distribution, and $\mathcal{F}^{(1)}_{gg}$, $\mathcal{F}^{(2)}_{qg}$ and $\mathcal{F}^{(3)}_{gg}$ are the additional distributions that enter two-particle production in forward $pA$ collisions~\cite{Dominguez:2011wm}. All distributions are expressed through the dipole in the Gaussian approximation at large $N_c$, and the dipole is evolved by the running-coupling BK equation~\cite{Balitsky:2006wa} from an MV initial condition at $x_0=0.01$ fitted to HERA data ($Q_{s0}^2=0.104$~GeV$^2$, $\Lambda=0.241$~GeV for the proton), with $Q^2_{s0,{\rm Pb}}=2.75\,Q^2_{s0,p}$~\cite{Deganutti:2023qct}. Bottom: ratio per unit transverse area normalized by the ratio of the initial saturation scales, $R_{{\rm Pb}/p}=\big[(\mathcal{F}/S_\perp)_{\rm Pb}/(\mathcal{F}/S_\perp)_{p}\big]\,Q^2_{s0,p}/Q^2_{s0,{\rm Pb}}$, which coincides with the ratio per nucleon when $S_{\perp,{\rm Pb}}Q^2_{s0,{\rm Pb}}=A\,S_{\perp,p}Q^2_{s0,p}$. The Cronin peak of the initial condition is washed out by evolution and replaced by a suppression below unity.}
\label{fig:gluonBK}
\end{figure}

The main qualitative effect of small-$x$ evolution is to increase the saturation scale $Q_s(x)$ as $x$ decreases, thereby shifting the transition between the dilute and saturated regimes towards larger transverse momenta. As previously discussed, the TMD is characterized by a dilute tail at $k_\perp\gg Q_s(x)$ and by a saturated regime at $k_\perp\lesssim Q_s(x)$. As a given process probes smaller and smaller $x$ values in the target, gluon radiation populates additional modes in the target wave function. Once the occupation of the low-$k_\perp$ modes becomes large, non-linear effects limit further growth in this region and the newly generated gluons are increasingly distributed towards larger transverse momenta. In this sense, small-$x$ evolution can be viewed as a broadening of the transverse-momentum distribution together with a growth of the saturation scale. Figure~\ref{fig:gluonBK} shows this behavior for the gluon TMDs of Sec.~\ref{sec:tmds}, namely the WW and dipole distributions and those that enter two-particle production in forward $pA$ collisions, with the dipole evolved by the running-coupling BK equation from an MV initial condition fitted to HERA data. The distributions differ markedly at $k_\perp\lesssim Q_s$, where the dipole and adjoint dipole distributions vanish as $k_\perp^2$, the WW distribution grows logarithmically and the others approach a plateau. All of them share a common perturbative tail at $k_\perp\gg Q_s$, and the turnover moves to larger $k_\perp$ as $x$ decreases. Compared with the $1/k_\perp^2$ tail of the initial condition (see also Table~\ref{tab:TMD-MV-asymptotics}), high-energy evolution modifies the power-law index at large $k_\perp$, generating an anomalous dimension governed by the BFKL kernel~\cite{Iancu:2002tr,Mueller:2002zm,Munier:2003vc,Munier:2003sj}.

It is also insightful to understand the impact of this evolution in the nuclear-to-proton ratio of the gluon TMDs, normalized by the number of nucleons. In the initial condition to the BK/JIMWLK equation, the larger saturation momentum of the nucleus leads to a characteristic Cronin enhancement in the $k_\perp$ domain where the proton is still dilute while the nucleus is already saturated. This enhancement is a property of the initial, quasi-classical distribution such as the MV model. Quantum evolution towards smaller $x$ values progressively washes out the Cronin peak because the proton and nuclear TMDs do not evolve at the same rate. The proton distribution in the dilute regime grows more rapidly with decreasing $x$, while the nuclear distribution, being already in the saturated regime over a larger range of transverse momenta, evolves more slowly. As a result, the initial enhancement of the nuclear-to-proton ratio is progressively reduced: the proton distribution catches up with, and eventually overtakes, the nuclear distribution in the region where the Cronin peak was initially present. 
The Cronin peak is thus washed out by small-$x$ evolution and replaced by a ratio that is suppressed below unity over a broad range of transverse momenta around the nuclear saturation scale, as shown in the bottom panels of Fig.~\ref{fig:gluonBK}, where the peak present at $x=10^{-2}$ has disappeared by $x=10^{-4}$.

\begin{figure}[t]
\centering
\includegraphics[width=\textwidth]{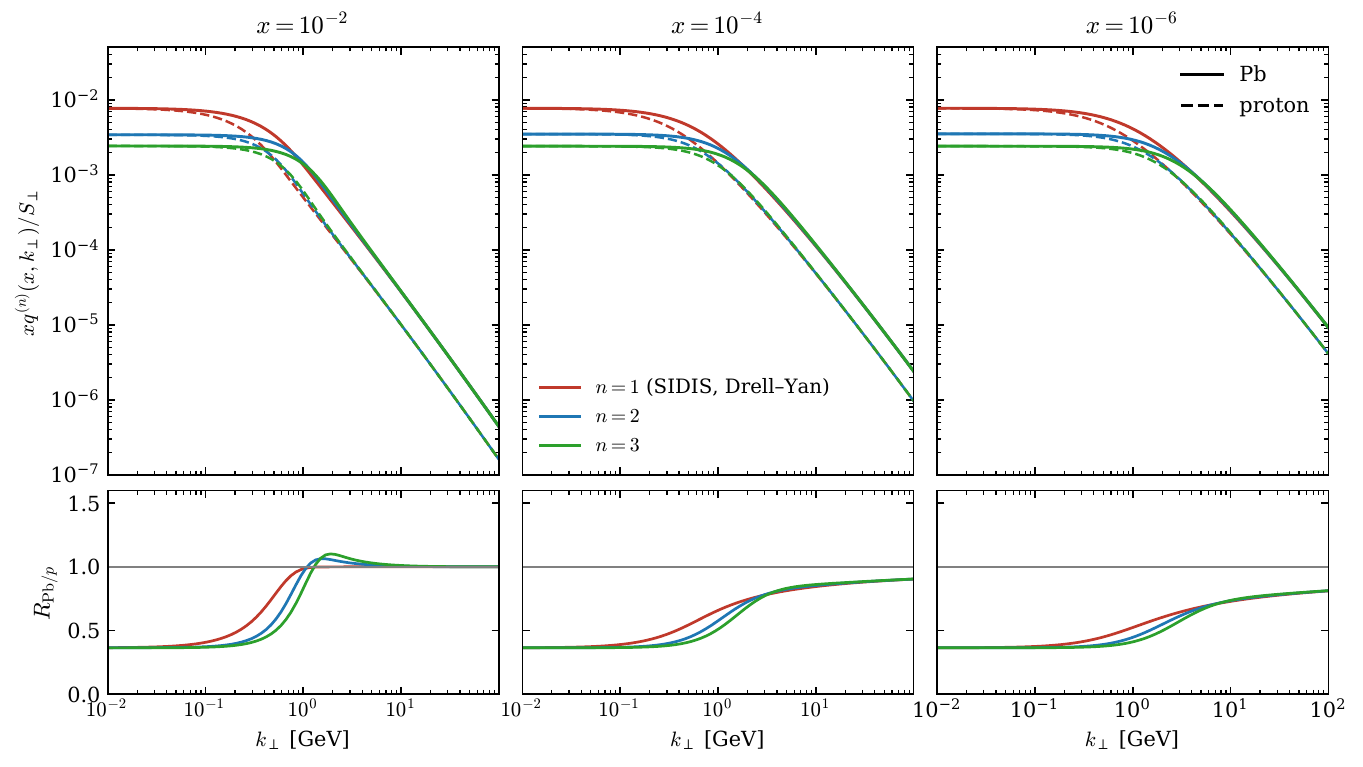}
\caption{Small-$x$ sea-quark TMDs $xq^{(n)}$, normalized by the transverse area $S_\perp$, for a lead nucleus (solid) and a proton (dashed) at $x=10^{-2}$, $10^{-4}$ and $10^{-6}$, obtained from the same evolved dipole as in Fig.~\ref{fig:gluonBK}. The distribution with $n=1$ is the one probed in SIDIS and Drell--Yan, while at large $N_c$ the distributions with $n=2,3$, which enter forward photon--jet and quark--gluon dijet production in $pA$ collisions, are obtained by convoluting it with $n-1$ dipole correlators~\cite{Caucal:2025xxh}. Bottom: nucleus-to-proton ratio $R_{{\rm Pb}/p}$, defined as in Fig.~\ref{fig:gluonBK}. The Cronin-like enhancement of the distributions with $n=2,3$ at the initial condition is washed out by evolution.}
\label{fig:seaquark}
\end{figure}

Sea-quark TMDs inherit this behavior, since their transverse-momentum distributions are generated from, and closely follow, the gluon distributions (Fig.~\ref{fig:seaquark}). At large $N_c$, the sea-quark distributions that enter forward $pA$ processes contain additional dipole correlators compared with the one probed in SIDIS~\cite{Caucal:2025xxh}. At the initial condition they display a Cronin-like enhancement in the nucleus-to-proton ratio that is absent for the SIDIS distribution, and evolution washes it out.

\subsection{Hard scale dependence: Sudakov suppression and Collins-Soper-Sterman evolution}
\label{sec:sudakov}

We recall that this review focuses on TMD parton distributions relevant for processes characterized by the hierarchy of scales $k_\perp^2 \ll Q^2 \ll s.$
The hierarchy $Q^2 \ll s$, or equivalently $x \ll 1$, provides a large logarithmic phase space for gluon radiation and gives rise to large $\mathcal{O}(\alpha_s \ln(1/x))$ corrections, which are resummed to all orders by the BK/JIMWLK equation, which we have just introduced. The other hierarchy of scales, $k_\perp^2 \ll Q^2$, is responsible for a different class of large corrections that arise when the process is computed at NLO in the small-$x$ regime: the Sudakov logarithms. These logarithms are typically double $\propto \ln^2(Q^2/k_\perp^2)$ and single logarithms $\propto \ln(Q^2/k_\perp^2)$ of the parametrically large ratio $Q^2/k_\perp^2\gg 1$. Thus, in the regime where TMD factorisation emerges at leading order, one also encounters large NLO corrections that must be resummed to all orders in order to obtain reliable perturbative predictions.

\subsubsection{Interplay between Sudakov and small-$x$ logarithms in fixed order calculations}

The possible overlap in phase space between the gluons contributing to BK/JIMWLK evolution and those contributing to Sudakov resummation, as well as the connection between the Sudakov logarithms emerging in higher-order CGC calculations and the Collins--Soper--Sterman (CSS) formalism at moderate $x$, have been the subject of intense investigation and debate over the past few years. A variety of approaches have been developed to address these questions. In~\cite{Mueller:2012uf,Mueller:2013wwa}, it was shown for the first time, from a small-$x$ calculation, that such Sudakov logarithms indeed arise in the NLO corrections to various processes exhibiting TMD factorisation at leading order, with a detailed calculation of small transverse momentum Higgs production in forward $pA$ collisions. These studies also demonstrated that a simultaneous small-$x$ and Sudakov resummation is possible~\cite{Xiao:2017yya}, with the phase spaces associated with the two types of logarithmic contributions being well separated, at least for the phase space contributing to the double logarithms. Thanks to the development of computational techniques for calculating NLO corrections to processes in dilute-dense collisions, the systematic separation of these contributions at NLO in the CGC, and its relation to the known perturbative instabilities of the NLL BK/JIMWLK evolution equations, have since been clarified~\cite{Taels:2022tza,Caucal:2022ulg,Caucal:2023nci}. The separation has also been demonstrated to single-logarithmic accuracy for a few processes, such as inclusive back-to-back dijet production in DIS~\cite{Caucal:2023nci,Caucal:2023fsf} and single-inclusive jet production~\cite{Caucal:2024vbv}. This implies, in particular, that the single Sudakov logarithms are not sensitive to the high-energy factorisation scale $k_f^-$. The bottom line is that a naive factorisation in $k_g^+$, obtained by introducing an arbitrary factorisation scale $k_f^+$ in Eq.\,\eqref{eq:slow-div} such that the phase space $k_g^+ \leq k_f^+$ is absorbed into the high-energy BK-JIMWLK evolution, while the phase space above $k_f^+$ is retained in the NLO coefficient function, does not provide a proper separation from the Sudakov double logarithms. As a consequence, applying such a naive factorisation scheme to e.g.~inclusive dijet production leads to a Sudakov double logarithm with the wrong sign, since part of the phase space associated with the Sudakov contribution is effectively absorbed into the high-energy evolution.

Physically, the naive factorisation scheme that we have just sketched amounts to ordering gluon emissions through the evolution in $k_g^+$, with $q^+ \gg k_{g,1}^+ \gg k_{g,2}^+ \gg \ldots$, where $q^+$ is the large momentum component of the dilute probe. A more physical choice, however, would be to order the gluon emissions according to their lifetimes $\tau_g\sim 1/k_g^-$, or equivalently according to decreasing $x$ values $P^- \gg \ldots \gg k_{g,2}^- \gg k_{g,1}^- \gg q^-$. In this picture, the BK/JIMWLK evolution is genuinely an evolution in $x = k_g^-/P^-$, with the first NLO gluon in the strict NLO result corresponding to the smallest $x$ value when viewed as part of the target wavefunction. It turns out that lifetime, or equivalently $k_g^-$, ordering is also essential for stabilising the NLL BK/JIMWLK equation~\cite{Ducloue:2019ezk} (and the BFKL equation in the dilute regime~\cite{Andersson:1995ju,Kwiecinski:1996td,Kwiecinski:1997ee,Salam:1998tj,Ciafaloni:1998iv,Ciafaloni:1999yw,Ciafaloni:2003rd,SabioVera:2005tiv,Motyka:2009gi}). Indeed, its NLL kernel contains large collinear double logarithms whose origin can be traced back to an incorrect identification of the ordering variable. Once lifetime ordering is effectively imposed, either through a modification of the LO kernel that resums these double collinear logarithms to all orders, or through a scheme transformation that allows one to change from one ordering variable to the other and thereby exactly cancel the large double collinear logarithms in the NLL BK/JIMWLK equation, the NLL equation becomes perturbatively reliable again. Changing the factorisation scheme not only modifies the NLL kernel of the high-energy evolution, but also the NLO coefficient function. Going from $k_g^+$ to $k_g^-$ factorisation therefore provides precisely the counterterm required in the NLO coefficient function to obtain the correct physical Sudakov double logarithm in small-$x$ calculations.

\begin{figure}[t]
\centering
\includegraphics[width=0.72\textwidth]{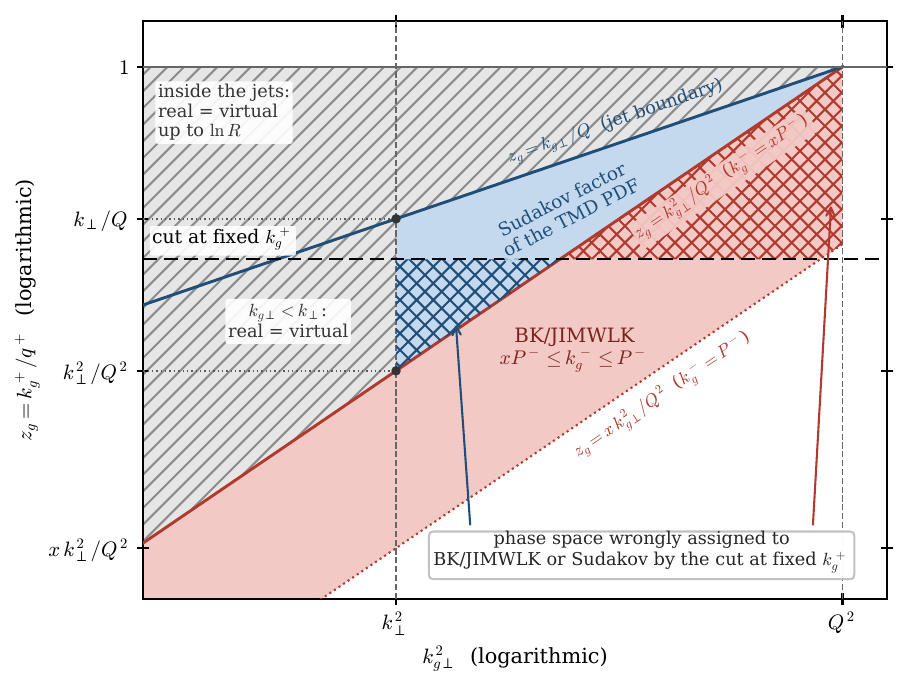}
\caption{Phase space of the NLO gluon in back-to-back dijet production in DIS, in terms of its transverse momentum $k_{g\perp}$ and momentum fraction $z_g=k_g^+/q^+$. Red: small-$x$ evolution of the target, below the line $\tau_g=\tau_\gamma\sim 1/q^-$, that is $z_g=k_{g\perp}^2/Q^2$ ($k_g^-=xP^-$), and above the dotted line $k_g^-=P^-$. Blue: Sudakov factor of the TMD PDF. Hatched: real and virtual emissions cancel. The horizontal dashed line is a cut at fixed $k_g^+$, which, for naive high-energy factorization along the projectile light-cone momentum, marks the separation between evolution and NLO coefficient-function.}
\label{fig:phasespace}
\end{figure}

Fig.~\ref{fig:phasespace} illustrates this separation of the gluon phase space for back-to-back dijets in DIS. The line $z_g=k_{g\perp}^2/Q^2$ marks where the lifetime of the gluon equals that of the photon fluctuation. Gluons with shorter lifetimes, below the line, are part of the evolution of the target, and the logarithm $\ln(1/x)$ is the height of this red band. Gluons with longer lifetimes are final-state radiation. Those emitted outside the jets with $k_{g\perp}>k_\perp$ form the blue triangle, whose area gives the Sudakov double logarithm. A factorization cut at fixed $k_g^+$ ignores the lifetime ordering and would instead assign part of this triangle to the evolution, which is the origin of the wrong sign discussed above. Emissions with $k_{g\perp}<k_\perp$ or inside the jets cancel between real and virtual contributions, up to the single logarithm of the jet radius.

On the practical side, several techniques can be employed to cleanly isolate the high-energy evolution from the NLO Sudakov logarithms to a TMD-factorisable process at small $x$: (i) implement a kinematic constraint in the $k_g^+$ integral of Eq.\,\eqref{eq:slow-div} that mimics the lifetime-ordering constraint $q^- \leq k_g^-$ for the radiated gluon at NLO. This is the strategy adopted in~\cite{Taels:2022tza,Caucal:2022ulg,Caucal:2023nci,Caucal:2023fsf}, and it should naturally be used in combination with the kinematically constrained or collinearly improved BK equation discussed in~\cite{Beuf:2014uia,Iancu:2015vea,Iancu:2015joa}. (ii) Alternatively, one can use naive $k_g^+$ factorisation and then implement a scheme transformation from $k_g^+$ ordering to $k_g^-$ ordering, following~\cite{Boussarie:2025mzh,Boussarie:2025bpq}. This strategy also naturally leads to a stable NLL BK equation with a modified NLO kernel free of double collinear logarithms. (iii) One can also employ a different regulator for the gluon phase-space integral, as in~\cite{Altinoluk:2025tms}, which properly isolates the appropriate phase space to be included in the high-energy evolution. The resulting evolution should then be equivalent, up to finite NLO corrections to the NLL kernel, to the NLL evolution equation obtained after the scheme transformation in strategy (ii). Each strategy has its pros and cons: strategy (i) is not as systematic as the other two, insofar as the kinematic constraint to be implemented may appear to be process dependent, since the condition $q^-\le k_g^-$ depends on the kinematics of the probe. Strategy (iii) may be difficult to implement in practice, since a new regulator mixing $k_g^+$ with the transverse momentum of the NLO gluon can lead to more complicated transverse integrals that may not be analytically tractable. Finally, strategy (ii) is still at a preliminary stage, as the general procedure for constructing a scheme transformation in the non-linear regime has so far only been established in the large-$N_c$ limit.

Irrespective of the computational strategy, the end result can be written schematically as
\begin{align}
    \der\sigma &= H_{\rm LO}(Q)\!\times\!\left[1-\alpha_sC_{\rm DL}\ln^2(Q^2/k_\perp^2)-\alpha_sC_{\rm SL}\ln(Q^2/k_\perp^2)+\alpha_s\ln(x_g/x_f)\mathcal{H}_{\rm JIMWLK}\otimes\right]F(x_f,\kt)\nonumber\\
&\quad+\alpha_s\times \textrm{``finite"}\,,\label{eq:sudakov-logs}
\end{align}
up to power corrections in $k_\perp/Q$, as usual. Here $C_{\rm DL}$ and $C_{\rm SL}$ are process-dependent pure numbers, independent of $k_f^-$. The ``finite'' term is an $\alpha_s$ correction with neither small-$x$ nor $Q/k_\perp$ logarithms. The last term in the square bracket, proportional to $\ln(x_g/x_f)$ is also a finite term provided $x_f\equiv k_f^-/P^-\sim x_g$ associated with the left-over after factorizing the divergence in $k_g^+$ in Eq.\,\eqref{eq:slow-div} with physical lifetime ordering. It is kept explicit in Eq.\,\eqref{eq:sudakov-logs} to highlight the structure of the factorization. Fixed-order CGC calculations at one loop featuring this structure are available for inclusive dijet production in DIS~\cite{Caucal:2023fsf} and in the photoproduction limit~\cite{Taels:2022tza}, as well as for single-inclusive hadron production in DIS~\cite{Altinoluk:2024vgg}, although the single-logarithmic coefficient $C_{\rm SL}$ remains to be computed in the latter case. The next frontier is twofold: (i) to extend these results to NLO for all phenomenologically relevant TMD-factorizable processes, and (ii) to obtain the first two-loop result, which would incorporate the NLL BK/JIMWLK Hamiltonian~\cite{Balitsky:2007feb,Balitsky:2009xg,Kovner:2013ona,Kovner:2014xia,Kovner:2014lca,Lublinsky:2016meo,Lappi:2020srm,Dai:2022imf} and subleading Sudakov logarithms of order $\mathcal{O}\bigl(\alpha_s^2\ln(Q^2/k_\perp^2)\bigr)$.

\subsubsection{Resummation of Sudakov logarithms}
\label{subsub:Sudakovresum}

Now that the Sudakov logarithms have been properly identified in fixed-order truncated perturbation theory, an important question is how to resum them to all orders. Although this generically remains an open problem, in particular for gluon TMDs with gauge-link structures more complicated than those of the WW TMD, there is increasing evidence that the resummation can be performed using an evolution equation formally equivalent to the Collins-Soper-Sterman equation. This has been shown for several processes and TMDs at small $x$, at least in the double-logarithmic approximation, where one only aims at resumming the terms of the form $\ln^2(Q^2 b_\perp^2)$ or $\ln^2(Q^2/k_\perp^2)$ to all orders, with $b_\perp$ the Fourier-conjugate variable to $k_\perp$. (This is in fact the natural variable that appears in NLO cross sections, as in~Eq.\,\eqref{eq:sudakov-logs}.) We review below the approaches developed for the resummation of Sudakov logarithms at small $x$.

\paragraph{Rapidity-only factorization}

At the operator level, the emergence of the CSS equation at double-logarithmic accuracy was demonstrated for the WW gluon TMD in~\cite{Balitsky:2015qba,Balitsky:2016dgz}, using rapidity-only factorization~\cite{Balitsky:2023hmh}. Note that, in this approach, the two WW gluon TMDs with staple-shaped gauge links pointing to the future~\cite{Balitsky:2015qba} and to the past~\cite{Balitsky:2016dgz}, relevant for back-to-back dijet production in DIS and Higgs production in $pA$ collisions, respectively, obey different rapidity evolution equations. Nevertheless, both equations reproduce the same physically relevant limits, namely the small-$x$, Dokshitzer--Gribov--Lipatov--Altarelli--Parisi (DGLAP), and CSS evolution equations in their respective regimes of validity. Working at the operator level (without referring to a particular process where the TMD operator appears) has the advantage that the rapidity divergences associated with the Sudakov logarithms appear explicitly. These divergences are then cancelled by renormalization of the operator. In the rapidity-only framework, the fields are decomposed into rapidity regions associated with the projectile, target, and central sectors. Quantum corrections to the target TMD operator are obtained by integrating out the central fields while keeping the projectile and target field configurations fixed. The resulting evolution equation, which depends on the rapidity divider separating the target fields from the other modes, is rather complicated. However, it admits simpler limits in specific kinematic regimes. As discussed in the section on high-energy evolution, varying the arbitrary rapidity divider leads to JIMWLK evolution of the WW gluon TMD when the transverse momenta involved are of comparable magnitude and much smaller than $\sqrt{s}$. Note that in the rapidity-only approach, this "external" rapidity divider is a priori independent of the $x$ argument of the TMD (see, in contrast, the next paragraph). On the other hand, in the regime where the rapidity divider is taken sufficiently large compared with $\sim k_\perp^2/Q^2$, the evolution equation becomes linear and reduces to the CSS equation in the double-logarithmic approximation. To some extent, the rapidity-only framework thus provides a unified approach to resumming both small-$x$ and Sudakov logarithms, much like the Ciafaloni-Catani-Fiorani-Marchesini (CCFM) equation~\cite{Ciafaloni:1987ur,Catani:1989sg,Marchesini:1994wr} in the linear regime (see below).

Several further developments of the rapidity-only approach have been pursued. The conformal properties of the evolution equation were investigated in~\cite{Balitsky:2019ayf}, while the implementation of running-coupling corrections was discussed in~\cite{Balitsky:2022vnb}. A simplified version of this equation, which incorporates both BFKL and Sudakov resummation, has also been proposed in a form amenable to numerical implementation~\cite{Balitsky:2026lop}. Related proposals can be found in Refs.~\cite{Duan:2024qck,Duan:2024qev}, where a generalized evolution equation for the WW gluon TMD is derived using the Born–Oppenheimer renormalization group approach. As in rapidity-only factorization, the evolution is formulated in terms of a single projectile energy-resolution scale, $k^+$, following the convention adopted in this review. This scale separates the modes in the projectile wave function that are averaged over in the TMD operator from those that are resolved in a given process involving the TMD. In the linear regime, this equation reduces to the CSS equation, while it also incorporates non-linear corrections, accounting for stimulated emission into highly occupied gluon modes, warrant further investigation. Another goal of these studies is to understand the transition from small to moderate $x$, where the collinear matching of the operator should connect to the standard PDF governed by DGLAP evolution. In this direction, a new TMD factorization scheme was developed in~\cite{Mukherjee:2023snp}, building on the rapidity-only factorization approach and background-field methods to separate collinear, anti-collinear, and soft modes. This framework makes it possible to go beyond the collinear limit, $k_\perp\gg Q_s,\Lambda_{\rm QCD}$, and to capture BFKL dynamics in the small-$x$ regime. Extending this TMD factorization scheme to the non-linear regime remains an open task.

\paragraph{Joint JIMWLK plus CSS evolution}

In contrast with the rapidity-only approach, whose evolution equation depends only on one rapidity scale and smoothly interpolates the high-energy JIMWLK evolution and the CSS evolution in the regime where the associated logarithms become large, a simultaneous resummation of high-energy logarithm and Sudakov logarithms is also possible~\cite{Marzani:2015oyb,Xiao:2017yya} by formulating their respective evolution in terms of two independent variables: (i) as discussed in section~\ref{sec:jimwlk}, small-$x$ logarithms are resummed via an evolution equation in $k_f^-=x_f P^-$, where $x_f$ then sets the $x$--scale of the TMD $x_f G_{\rm WW}(x_f,\kt)$ with typically (to minimize the NLO impact factor) $x_f=x_g$ with $x_g$ the target longitudinal momentum fraction of the small-$x$ gluons probed by the process, (ii) the Sudakov logarithms are resummed via an evolution equation formulated in terms of the hard scale $Q$ of the process, such that the TMD acquires an \textit{additional} scale dependence $x_f G_{\rm WW}(x_f,\kt)\to x_f G_{\rm WW}(x_f,\kt; Q^2)$. The connection with the CSS formalism can be made more manifest by introducing in fact two additional scales, one UV scale $\mu^2$ associated with standard UV renormalization and the Collins scale $\zeta=2(xP^-)^2 e^{2y_{\rm max}}$ associated with the rapidity cut-off $y_{\rm max}$ of the gluon which contributes to the CSS evolution, while not being part of the BK/JIMWLK evolution. 

This joint small-$x$ and Sudakov resummation approach can be demonstrated at the operator level~\cite{Xiao:2017yya} or by computing the NLO corrections to a physical cross section. In this context, as in Eq.\,\eqref{eq:sudakov-logs}, the Sudakov logarithms do not give rise to an actual divergence, since they are always regulated by external physical scales associated with the process under consideration. The CSS equation for the WW gluon TMD was established in~\cite{Caucal:2024bae,Caucal:2025mth} following this strategy, by studying the dependence of the one-loop correction to inclusive back-to-back dijet production in DIS on the rapidity scale separating gluon emissions inside and outside the jets. (Gluon emissions inside the jets would instead contribute to the evolution of the jet function in TMD factorization theorems involving jets~\cite{Gutierrez-Reyes:2018qez,Gutierrez-Reyes:2019vbx,delCastillo:2020omr,delCastillo:2021znl}.)

In this calculation, the divergence as the projectile light-cone momentum of the radiated gluon goes to zero ($k_g^+=z_gq^+\to 0$) in the NLO cross section has already been cancelled by the high-energy renormalization of the bare TMD. When formulated as an evolution equation in $k_g^-$, this renormalization effectively amounts to absorbing into the evolution the phase space for the NLO gluon defined by $x k_{g\perp}^2/Q^2\le z_g\lesssim k_{g\perp}^2/Q^2$, without any constraint on $k_{g\perp}$. Equivalently, the absorbed phase space corresponds to $x P^-\le k_g^-\lesssim P^-$ that gives the $\ln(1/x)$ enhancement of the NLO cross-section. Nevertheless, a logarithmic sensitivity to the upper cutoff of the $k_g^+$ integral remains, with the constraint $z_g\le k_{g\perp}/Q$ arising from the physical requirement that the real NLO gluon lie outside the jets formed by the quark and antiquark. This $Q$-dependent scale therefore provides the natural physical cutoff for the rapidity divergence, in the CSS terminology, in this particular process. Indeed, setting $\zeta=Q^2$ in the definition of the rapidity cutoff $y_{\rm max}$ such that $\zeta=2(xP^-)^2e^{2y_{\rm max}}$, yields precisely the condition $z_g\lesssim k_{g\perp}/Q$. The real correction also contains a contribution that, in the limit $k_\perp\gg Q_s$, reduces to a single step of the DGLAP evolution of the gluon PDF~\cite{Caucal:2024bae}. In this limit, the $k_\perp$ dependence of the real cross section is controlled by the recoil against the radiated gluon. This contribution is precisely what is required to ensure consistent collinear matching of the TMD at large $k_\perp$ to a DGLAP-evolved gluon PDF.

For generic values of $k_\perp$, including in the saturation regime~\cite{Caucal:2025mth}, the $Q^2$ dependence of the WW gluon TMD as obtained in the CGC effective theory reads, to leading logarithmic accuracy
\begin{align}
    \frac{\partial xG_{\rm WW}(x,\boldsymbol{k};Q^2)}{\partial\ln Q^2}&=\frac{\alpha_sN_c}{2\pi^2}\int\frac{\der^2\ellt}{\ellt^2}\left[xG_{\rm WW}(x,\boldsymbol{k}+\ellt;Q^2)-\Theta\left(Q-|\ellt|\right)xG_{\rm WW}(x,\boldsymbol{k};Q^2)\right]\nonumber\\
&\quad+\frac{\alpha_sN_c}{\pi}\beta_0xG_{\rm WW}(x,\boldsymbol{k};Q^2)\,. \label{eq:CSS-ms}
\end{align}
A similar equation has been recently derived for the unintegrated gluon distribution in $k_t$ factorization~\cite{vanHameren:2025hyo} (the connection is manifest in the $x\to0$ limit). In the conformal case ($\beta_0=0$), this equation admits a clear physical interpretation as a \textit{rate equation}. As the phase space for soft gluon radiation increases, it describes the splitting of a gluon with transverse momentum $\kt+\ellt$ into a soft, unmeasured gluon with transverse momentum $\ellt$, characterized by the typical Bremsstrahlung behavior $\propto 1/\ellt^2$, and a gluon that contributes to the $\kt$ bin of the gluon TMD distribution. This real-emission process is balanced by the virtual term, but only below the resolution scale $Q$. This ensures the conservation of the total number of gluons with transverse momentum below $Q$ during the evolution. Since the unmeasured gluon carries a very small target longitudinal momentum fraction, the equation is local in $x$ as emitting this gluon does not degrade the $x$ value of its parent. The $\beta_0$ term in Eq.\,\eqref{eq:CSS-ms}, with $\beta_0=(11N_c-2N_f)/(12N_c)$, is of UV origin, and arises in the CGC framework from the one-loop correction to the classical background field~\cite{Ayala:1995hx,Gelis:2006yv,Gelis:2006cr}. The UV divergence of this contribution is removed by standard renormalization, which amounts to replacing the bare coupling in Eq.\,\eqref{eq:WW-TMD} for the WW gluon TMD by the renormalized one $\alpha_s(\mu)$. The running of $\alpha_s$ from the scale $\sim k_\perp$ to the hard scale $Q$ of the process generates a Sudakov single logarithm captured by the $\beta_0$ dependent term in Eq.\,\eqref{eq:CSS-ms}. This result is confirmed by the explicit computation of the anomalous dimension of the WW gluon TMD in the saturation regime done in~\cite{Zhou:2018lfq}.

Eq.\,\eqref{eq:CSS-ms} turns out to be equivalent to the soft $z \to 1$ limit of the CCFM equation (see also earlier seminal studies~\cite{Parisi:1979se,Kodaira:1981nh}). Furthermore, Eq.\,\eqref{eq:CSS-ms} is easily solved in coordinate space, where it is shown to be equivalent to the ``diagonal'' version of the two-scale CSS equations where the Collins-Soper scale $\zeta$ and the UV renormalization scale $\mu^2$ are identified and both set to the hard scale $Q^2$ at the end of the evolution~\cite{Caucal:2024bae}. Therefore, in coordinate space, the resummed result for a small-$x$ WW gluon TMD takes the standard exponential Sudakov suppression form, through which the TMD acquires an explicit dependence on the hard scale through CSS evolution, of the form
\begin{align}
x_fG_{\rm WW}(x_f,\kt; Q^2)&=\int\der^2\boldsymbol{b}_\perp \ e^{i\kt\cdot\boldsymbol{b}_\perp}\exp\left(-\int_{\mu_b^2}^{Q^2}\frac{\der\mu^2}{\mu^2}\left[A_i\ln\left(\frac{Q^2}{\mu^2}\right)+B_i\right]\right)\nonumber\\
&\qquad\times x_f\tilde{G}_{\rm WW}(x_f,\boldsymbol{b}_\perp;\mu_b^2)\,,\label{eq:sudakov}
\end{align}
with $\mu_b=c_0/b_\perp$ and $c_0=2e^{-\gamma_E}$. This choice of $\mu_b$ comes naturally from the Fourier transform of Eq.\,\eqref{eq:CSS-ms}. It is also the value that emerges directly from the Sudakov logarithms in NLO calculations~\cite{Mueller:2013wwa,Taels:2022tza,Caucal:2022ulg}, in particular when deriving the correct single-logarithmic coefficient $C_{\rm SL}$, including in the dense regime $k_\perp\lesssim Q_s$. To first order in $\alpha_s$, the Sudakov factor of Eq.\,\eqref{eq:sudakov} reduces to $1-\frac{A_i}{2}\ln^2(Q^2/\mu_b^2)-B_i\ln(Q^2/\mu_b^2)$, which reproduces the contribution of the TMD to the coefficients $C_{\rm DL}$ and $C_{\rm SL}$ of Eq.\,\eqref{eq:sudakov-logs}, with $k_\perp$ replaced by $\mu_b$.
In Eq.\,\eqref{eq:sudakov}, $x\tilde{G}_{\rm WW}(x,\boldsymbol{b}_\perp;\mu_b^2)$ is the Fourier transform\footnote{We use $x\tilde{G}_{\rm WW}(x,\boldsymbol{b}_\perp;\mu^2)=\int\frac{\der^2\kt}{(2\pi)^2}\,e^{-i\kt\cdot\boldsymbol{b}_\perp}\,xG_{\rm WW}(x,\kt;\mu^2)$, consistent with the inverse transform in Eq.~\eqref{eq:sudakov}.} of the small-$x$ evolved TMD obtained by solving the JIMWLK equation for a given initial condition at some initial scale $x_0$ down to the high-energy factorization scale $x_f\sim x_g\ll 1$. Note that while Eq.\,\eqref{eq:CSS-ms} and the CSS coordinate space version are equivalent, solving these two equations as boundary value problems in coordinate space $x\tilde{G}_{\rm WW}(x,\boldsymbol{b}_\perp;Q^2=\mu_b^2)$ or in momentum space $xG_{\rm WW}(x,\kt;Q^2=\kt^2)$ leads to different solutions. This aspect has been discussed for the quantum evolution of the diffractive gluon distribution introduced in Eq.\,\eqref{eq:gluon-diffTMD} in~\cite{Iancu:2025jsu}.

 The coefficients $A_i$ and $B_i$ depend on the flavour of the TMD. 
 For the WW gluon TMD in Eq.\,\eqref{eq:sudakov}, $A_g=\alpha_sN_c/(2\pi)$ and $B_g=-\alpha_sN_c\beta_0/\pi$, with $\beta_0$ as in Eq.~\eqref{eq:CSS-ms}, while for the sea-quark TMD, at one loop, $A_q=\alpha_sC_F/(2\pi)$ and $B_q=-3\alpha_sC_F/(4\pi)$. As already mentioned, Eq.\,\eqref{eq:sudakov} has been established at one loop for the WW gluon TMD  involved, for example, in inclusive dijet production in DIS~\cite{Balitsky:2015qba,Taels:2022tza,Caucal:2023fsf}. Note, however, the single-logarithmic coefficient $C_{\rm SL}$ as defined in Eq.\,\eqref{eq:sudakov-logs} receives contributions in addition to $B_g$. How these additional contributions should be resummed, and in particular whether they exponentiate in the same way as in the CSS solution, remains an open question (for the single logarithm depending on the jet radius $R$, the resummation should be implemented at the level of the jet function). Similar expressions as Eq.\,\eqref{eq:CSS-ms} or Eq.\,\eqref{eq:sudakov} hold for the (sea) quark TMD involved in single inclusive jet production in DIS, as shown in~\cite{Caucal:2024vbv} in the dilute limit $k_\perp\gg Q_s$ only, provided the jet is reconstructed with a suitable jet definition.  In the saturation regime $k_\perp\sim Q_s$, semi-inclusive jet or hadron production with $k_\perp^2\ll Q^2$ remains to be computed at NLO.

Recent results have also been obtained for processes involving TMD fragmentation functions, in particular for single inclusive hadron production~\cite{Altinoluk:2024vgg} and back-to-back dihadron production in DIS~\cite{Caucal:2024nsb}. In both cases, the factorisation formula involves a TMD fragmentation function whose associated Sudakov factor is also found to agree with the CSS solution. 

\begin{figure}[t]
\centering
\includegraphics[width=0.72\textwidth]{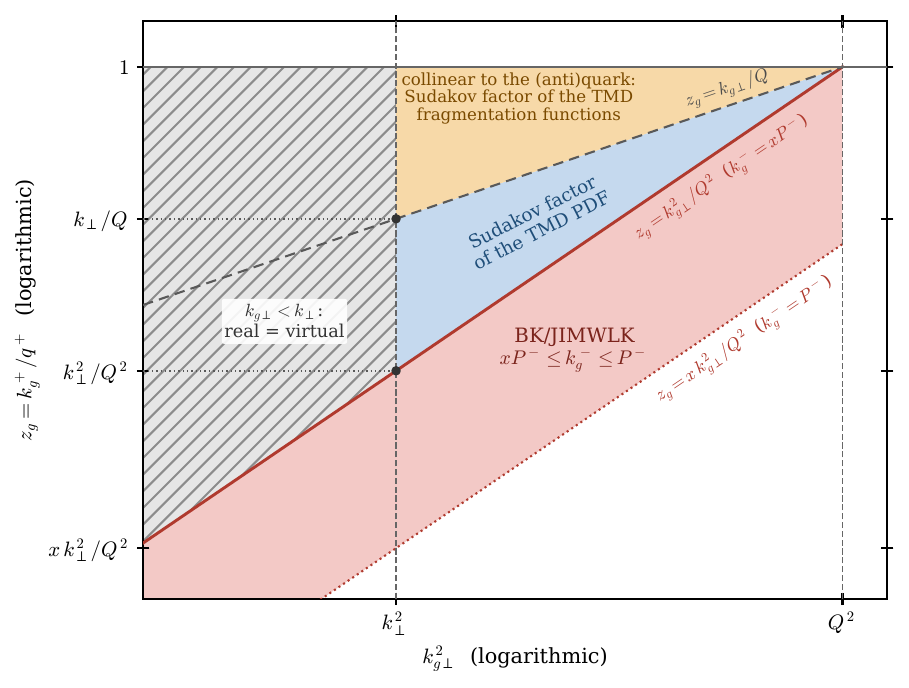}
\caption{Same as Fig.~\ref{fig:phasespace} for back-to-back dihadron production. Orange: the region inside the former jets, which is no longer cancelled for $k_{g\perp}>k_\perp$ and gives the Sudakov factor of the TMD fragmentation functions. For collinear emission $k_{g\perp}$ is measured relative to the radiating quark.}
\label{fig:phasespace-dihadron}
\end{figure}

For dihadrons, Fig.~\ref{fig:phasespace-dihadron} shows that the division of the phase space changes. Without a jet cone, a gluon collinear to the quark or antiquark changes the imbalance of the hadron pair, so that the region above $z_g=k_{g\perp}/Q$ is no longer cancelled for $k_{g\perp}>k_\perp$ (orange). Its collinear singularity is absorbed into the TMD fragmentation functions, which acquire their own Sudakov factor, and it accounts for the difference between the double-logarithm coefficients $-\alpha_sN_c/4\pi$ for dijets and $-(\alpha_s/2\pi)(C_F+N_c/2)$ for dihadrons~\cite{Caucal:2024nsb}. This is why the collinear fragmentation functions have to be promoted to TMD ones, in order to preserve the universality of the Sudakov soft factor of the gluon TMD.

For heavy quarks produced in the final state, a Sudakov resummation scheme has been proposed in~\cite{Marquet:2025jdr} to account for the additional quark mass scale, which introduces the large logarithm $\ln(Q^2/m^2)$ besides $\ln(Q^2/k_\perp^2)$.

For forward two-particle production in proton-nucleus collisions, it remains an open question whether TMD factorisation, supplemented by CSS resummation for each TMD distribution and fragmentation function, can be consistently matched to an NLO small-$x$ calculation. In such processes the soft factor is a matrix in color space, so that the Sudakov resummation mixes the different gluon TMDs rather than acting on each of them separately~\cite{Sun:2014gfa}. For the gluon dipole TMD at small $x$, whose gauge-link structure involves both initial- and final-state interactions, an equation analogous to Eq.\,\eqref{eq:sudakov} has been established, but only at double logarithmic accuracy~\cite{Mueller:2013wwa,Xiao:2017yya}. An ongoing NLO study of forward photon--jet production in the back-to-back limit suggests that, in the saturation regime, the CSS equation may receive non-linear corrections beyond the standard double-logarithmic exponentiation~\cite{Marcos-toappear}. Whether these corrections can be described by the CSS--JIMWLK Hamiltonian, which generates the evolution in rapidity at fixed $x$ of Wilson-line operators and reproduces the double-logarithmic CSS equation for the gluon TMD~\cite{Duan:2024nlr}, remains to be understood.

\paragraph{Phenomenological consequences of Sudakov suppression}

The effect of Sudakov resummation on the transverse-momentum dependence of the TMD is to induce an additional broadening of the distribution, which can be much larger than the broadening generated by saturation. This does not necessarily prevent the use of TMD observables to search for saturation effects at small $x$. Indeed, if sufficiently precise data and precise theoretical calculations are available, the Sudakov contribution can, at least in principle, be disentangled from the intrinsic transverse-momentum dependence of the TMD. One may then extract quantities characterising the saturation dynamics, such as the saturation scale $Q_s$, in much the same way as the non-perturbative part of a TMD is extracted from data in conventional TMD factorisation. In this sense, the Sudakov broadening does not erase the saturation signal. Rather, it provides an additional, calculable contribution that must be properly accounted for when extracting the underlying small-$x$ TMD. We shall return to the state of the art in the phenomenology of the joint small-$x$ plus CSS evolution for various processes in section~\ref{sec:pheno}.

\section{TMDs at small $x$ beyond the eikonal limit}
\label{sec:beyond-eikonal}

So far, we have mostly considered the strict $x\to 0$ limit of TMDs, i.e.~we have retained only the leading power in the $x$ expansion. At leading power in $x$, the only $x$ dependence arises from quantum corrections through the RG evolution described by the JIMWLK equation in Sec.~\ref{sec:jimwlk}. Recently, considerable efforts have been devoted to extending the CGC formalism by including so-called subeikonal corrections, or, more formally, power corrections in $x$~\cite{Altinoluk:2014oxa,Altinoluk:2015gia,Altinoluk:2015xuy,Altinoluk:2020oyd,Altinoluk:2021lvu,Altinoluk:2022jkk,Altinoluk:2023qfr,Agostini:2024xqs,Altinoluk:2024zom,Altinoluk:2024dba,Altinoluk:2024tyx,Agostini:2025vvx,Chirilli:2026pkv,Chirilli:2026vij,Chirilli:2021lif,Chirilli:2018kkw,Brown:2026qlc}.\footnote{This expansion, in inverse powers of the collision energy at fixed transverse momenta, is distinct from the expansion in $k_\perp/Q$ that underlies TMD factorization and from the saturation corrections in $Q_s/Q$ (Sec.~\ref{sec:itmd}). The two expansions nevertheless mix in the back-to-back limit, where next-to-eikonal corrections also generate twist-three TMDs~\cite{Altinoluk:2024zom}.} The motivation is twofold. (i) For TMDs that are not power suppressed in $x$, sub-eikonal corrections restore part of the $x$ dependence lost in the eikonal limit, such as the longitudinal phase of the TMD operators~\cite{Altinoluk:2024zom}, and provide a first step toward the moderate-$x$ regime. (ii) For TMDs that vanish at leading power in $x$, in particular several distributions of a polarized target, sub-eikonal accuracy is needed to access their small-$x$ behavior at all~\cite{Cougoulic:2022gbk}. By gaining greater theoretical control over the $x\to 0$ behaviour of such subeikonal TMDs, one hopes to better constrain the contribution of small-$x$ gluons and sea quarks to the total spin of the nucleon.

We will not review here the details of the formalism used to compute small-$x$ observables to subeikonal accuracy. The bottom line is that, in addition to the standard eikonal Wilson line dressing the quark and gluon fields, which accounts for their colour rotation as they propagate through the shockwave background field of the dense target, one must include subleading terms, with each term containing subeikonal operator insertions between segments of eikonal Wilson lines, suppressed by one power of the center-of-mass energy. For instance, the scattering amplitude of a fast quark with incoming helicity state $\sigma$ and outgoing state $\sigma'$ through the target, or equivalently, the parton propagator, can be organized as an expansion in which each sub-eikonal order carries a factor $P^-/s$, whose inverse dimension of energy is compensated by the integral along $z^+$ and the dimension-two operator insertions in Eq.~\eqref{eq:subeikonal-quark-Smatrix}, so that at fixed transverse momenta each order is suppressed by one power of $x$:
\begin{align}
U_{\boldsymbol{x},\boldsymbol{y};\sigma'\sigma}
&=
U(\boldsymbol{x})\delta^{(2)}(\boldsymbol{x}-\boldsymbol{y})
\delta_{\sigma'\sigma}
+
U^{(1)}_{\boldsymbol{x},\boldsymbol{y};\sigma'\sigma}
+
U^{(2)}_{\boldsymbol{x},\boldsymbol{y};\sigma'\sigma}
+\cdots \,,
\label{eq:subeikonal-expansion}
\end{align}
where the leading term is the eikonal Wilson line defined in Eq.\,\eqref{eq:Wilson-line} and the first subeikonal correction $U^{(1)}_{\boldsymbol{x},\boldsymbol{y};\sigma'\sigma}$ can be decomposed as
\begin{align}
&U^{(1)}_{\boldsymbol{x},\boldsymbol{y};\sigma'\sigma}
=\frac{iP^-}{s}
\int_{-\infty}^{\infty} \der z^+\,
\der^2\boldsymbol{z}\,
U_{\boldsymbol{x}}[\infty,z^+]\,
\delta^{(2)}(\boldsymbol{x}-\boldsymbol{z})
\nonumber\\
&\qquad\times\left[
-\delta_{\sigma\sigma'}\,
\overleftarrow{D}^{\,i}D^i
-g\,\sigma\,\delta_{\sigma\sigma'}\,F^{12}
\right](z^+,\boldsymbol{z})\,
U_{\boldsymbol{y}}[z^+,-\infty]\,
\delta^{(2)}(\boldsymbol{y}-\boldsymbol{z})
\nonumber\\
&\quad-\frac{g^2P^-}{2s}\,
\delta^{(2)}(\boldsymbol{x}-\boldsymbol{y})
\int_{-\infty}^{\infty} \der z_1^+
\int_{z_1^+}^{\infty} \der z_2^+\,
U_{\boldsymbol{x}}[\infty,z_2^+]\,
t^b\,\psi_\beta(z_2^+,\boldsymbol{x})\,
\mathcal{U}_{\boldsymbol{x}}^{ba}[z_2^+,z_1^+]
\nonumber\\
&\qquad\times\left[
\delta_{\sigma\sigma'}\,\gamma^-
-\sigma\,\delta_{\sigma\sigma'}\,\gamma^-\gamma^5
\right]_{\alpha\beta}
\bar{\psi}_\alpha(z_1^+,\boldsymbol{x})\,
t^a\,U_{\boldsymbol{x}}[z_1^+,-\infty]
+\ldots \,,
\label{eq:subeikonal-quark-Smatrix}
\end{align}
where our conventions differ from those of~\cite{Cougoulic:2022gbk} by the reflection $z\to-z$ (see footnote~\ref{fn:conventions}), and the $\ldots$ denote contributions from the finite longitudinal width of the shockwave, which cannot be neglected at sub-eikonal accuracy, and quark-mass corrections, which are usually neglected. Here $\sigma$ and $\sigma'$ are the light-cone helicities of the incoming and outgoing projectile quark, which moves along the light-cone plus direction. The complete sub-eikonal Wilson lines have been derived~\cite{Chirilli:2021lif,Altinoluk:2021lvu,Kovchegov:2021iyc,Cougoulic:2022gbk}. We denote by $U_{\boldsymbol{x}}[x_2^+,x_1^+]$ the finite Wilson-line segments at fixed transverse coordinate $\boldsymbol{x}$, and the corresponding adjoint Wilson line is denoted by $\mathcal{U}_{\boldsymbol{x}}[x_2^+,x_1^+]$, while $t^a$ are the generators in the fundamental representation and $\psi$ is the quark field of the target. $D^i$ is the covariant derivative, and in the helicity independent term, $\overleftarrow{D}^{\,i}$ is acting on the Wilson line on the left. The terms proportional to $\sigma$, namely the $F^{12}$ insertion and the axial quark current $\gamma^-\gamma^5$, are helicity dependent. The term proportional to $F^{12}$ physically represents the spin-chromomagnetic coupling similar to the QED case in atomic spin physics.

The helicity-independent term with $\overleftarrow{D}^{\,i}D^i$ in Eq.\,\eqref{eq:subeikonal-quark-Smatrix} has a simple physical origin: it accounts for the transverse motion of the parton inside the target, which is neglected at eikonal accuracy and enters at sub-eikonal order through its transverse kinetic energy, suppressed by the large longitudinal momentum. As part of the definition of the sub-eikonal Wilson line, this term must retain the full covariant derivatives. When the Wilson line enters a specific distribution, however, the transverse derivatives can be integrated by parts, and the term reduces to a simpler operator that depends on the distribution. For the gluon helicity TMDs discussed below, this operator is
\begin{align}
    U^{G[2],i}(\boldsymbol{x})&=\frac{P^-}{2s}\int_{-\infty}^{\infty}\der x^+\, U_{\boldsymbol{x}}[\infty,x^+]\left[D^i(x^+,\boldsymbol{x})-\overleftarrow{D}^{\,i}(x^+,\boldsymbol{x})\right]U_{\boldsymbol{x}}[x^+,-\infty] \,, \label{eq:subeikonal-UG2}
\end{align}
with $D^i=\partial^i-igA^i$ and $\overleftarrow{D}^{\,i}=\overleftarrow{\partial}^{\,i}+igA^i$ the right- and left-acting covariant derivatives in the fundamental representation.

Equations \eqref{eq:subeikonal-quark-Smatrix} and \eqref{eq:subeikonal-UG2} contain the main building blocks of the sub-eikonal helicity-dependent gluon TMDs at small $x$. High-energy evolution generates further operators beyond Eq.\,\eqref{eq:subeikonal-quark-Smatrix} and its ellipsis, such as Wilson lines that change from the fundamental to the adjoint representation, or vice versa, as they cross the shockwave~\cite{Borden:2024bxa}.

\subsection{Subeikonal spin-dependent TMDs}

Let us consider first the gluon TMDs for polarized targets. As discussed in Sec.~\ref{subsub:sivers}, the dipole-type gluon Sivers function, as well as the dipole-type $h_1$ and $h_{1T}^\perp$, do not vanish at leading power in $x$ and are given by the odderon operator, whereas their WW-type counterparts have no eikonal contribution. This is not the case for the gluon helicity TMD $g_{1L}$ which vanishes at eikonal accuracy.
Recently, there has been much progress in our understanding of the gluon helicity TMD $g_{1L}$ at small $x$. Starting from its operator definition~\cite{Kovchegov:2017lsr,Kovchegov:2018znm} or from an observable factorizing in terms of this TMD~\cite{Kovchegov:2015pbl}, one can obtain the CGC expression in terms of polarized Wilson line. 

The gluon helicity TMD with WW-type gauge link reads
\begin{align}
g_{1L}^{\rm WW}(x,\kt)&=\frac{2}{\alpha_s}\int\frac{\der^2\boldsymbol{x}\der^2\boldsymbol{y}}{(2\pi)^4}e^{-i\kt\cdot(\boldsymbol{x}-\boldsymbol{y})}\epsilon^{ij}\left\langle\left\langle \textrm{Tr}\left[ U(\boldsymbol{y})\partial^iU^\dagger(\boldsymbol{y})U^{G[2],j}(\boldsymbol{x})U^\dagger(\boldsymbol{x})\right]+c.c.\right\rangle\right\rangle_x \,. \label{eq:g1LWW}
\end{align}
As compared to the unpolarised WW gluon TMD defined in Eq.\,\eqref{eq:WW-TMD}, the double angle brackets introduced in~\cite{Kovchegov:2015pbl} scale out the explicit energy suppression of the sub-eikonal operator, $\langle\langle\ldots\rangle\rangle_x\equiv zs\,\langle\ldots\rangle_x$, with $z$ the fraction of the projectile momentum carried by the parton that crosses the shockwave, where for the helicity distributions the single brackets denote the difference between the two helicity states of the target, $\langle\ldots\rangle_x\equiv\frac{1}{2}\sum_{S_L}S_L\,\langle P,S_L|\ldots|P,S_L\rangle/(2P^-V^+)$, with $V^+$ the light-cone volume of the target~\cite{Kovchegov:2025gcg}. 

The WW distribution of Eq.\,\eqref{eq:g1LWW} is accessed through the longitudinal double-spin asymmetry of inclusive dijet production in polarized electron-proton collisions, which in the back-to-back limit probes it at leading power and without contamination from linearly polarized gluons~\cite{Kovchegov:2025gcg}. The dipole-type gluon helicity TMD reads
\begin{align}
    g_{1L}^{D}(x,\kt)&=-\frac{2i}{\alpha_s}\epsilon^{ij}\boldsymbol{k}_\perp^j\int\frac{\der^2\boldsymbol{x}\der^2\boldsymbol{y}}{(2\pi)^4}e^{-i\kt\cdot(\boldsymbol{x}-\boldsymbol{y})}\left\langle\left\langle \textrm{Tr}\left[U^{G[2],i}(\boldsymbol{x}) U^\dagger(\boldsymbol{y})\right]+c.c.\right\rangle\right\rangle_x \,.
\end{align}
These expressions are the analogs of Eqs.\,\eqref{eq:WW-TMD} and~\eqref{eq:dip-CGC}: they should be understood as ``bare" or tree-level expressions without quantum corrections. Unlike the unpolarized case, where the MV model provides a standard phenomenological initial condition for the small-$x$ evolution of Wilson-line correlators, whose parameters have been fitted to DIS and proton-nucleus data, the initial conditions for polarized Wilson-line operators remain less understood and less constrained. For helicity-dependent observables, a helicity-dependent extension of the MV model was first constructed in~\cite{Cougoulic:2020tbc} and completed in~\cite{Li:2024fdb,Li:2024xra}, in which the usual eikonal color sources are supplemented by subeikonal helicity-dependent sources. This model could provide an initial condition for small-$x$ helicity evolution. A helicity-dependent generalization of the JIMWLK equation is, however, not yet known, and even at large $N_c$ the non-linear terms enter at single-logarithmic accuracy, where linear contributions of the same order remain to be computed. Alternative initial conditions for the polarized dipole amplitudes have been obtained from explicit models of the proton wave function, in particular a three-valence-quark model~\cite{Dumitru:2024pcv}, providing a more direct connection between the polarized Wilson-line correlators and the proton's spin structure. 

The TMD $h_{1L}^\perp$, which can be accessed through $J/\psi$ production in electron scattering off a longitudinally polarized proton~\cite{liu:2023unl}, has recently been investigated at small $x$ in~\cite{Kovchegov:2025gcg}. Although its operator definition carries an eikonal prefactor, its bare contribution requires a classical gluon field $A^-$ in light-cone gauge that couples to the longitudinal polarization of the target, which the eikonal field does not, so that the WW-type distribution vanishes at eikonal order. The sub-eikonal classical field~\cite{Li:2024fdb,Li:2024xra} does not appear to couple to the corresponding operator either, and it has been suggested that the WW-type $h_{1L}^\perp$ may vanish at sub-eikonal order as well, which remains to be proven~\cite{Kovchegov:2025gcg}. To our knowledge, the remaining TMD $g_{1T}$, which is also expected to be at least subeikonal, has not yet been computed at small $x$.

\begin{table}[t]
    \centering
    \renewcommand{\arraystretch}{1.8}
    \setlength{\tabcolsep}{7pt}

    \begin{tabular}{
        >{\centering\arraybackslash}m{2.0cm}
        >{\centering\arraybackslash}m{3.7cm}
        >{\centering\arraybackslash}m{3.7cm}
        >{\centering\arraybackslash}m{3.7cm}
    }
        \rowcolor{gluonred}
        \multicolumn{4}{c}{
            \color{white}\large\bfseries
            Gluon TMD distributions
        } \\

        \rowcolor{gluonlight}
        \multicolumn{1}{c}{\textbf{Nucleon}}
        &
        \multicolumn{3}{c}{\textbf{Gluon operator polarization}} \\

        \rowcolor{gluonlight}
        \textbf{polarization}
        & \textbf{Unpolarized}
        & \textbf{Circular}
        & \textbf{Linear} \\

        \midrule

        \rowcolor{gluonrow}
        $\boldsymbol{U}$
        &
        $\displaystyle
        f_1 \sim \frac{1}{x}
        x^{-\frac{4\alpha_s N_c\ln 2}{\pi}}
        $
        &
        \textemdash
        &
        $\displaystyle
        h_1^\perp \sim \frac{1}{x}
        x^{-\frac{4\alpha_s N_c\ln 2}{\pi}}
        $
        \\

        \addlinespace[3pt]

        $\boldsymbol{L}$
        &
        \textemdash
        &
        $\displaystyle
        g_{1L} \sim
        x^0 x^{-3.66\sqrt{\frac{\alpha_s N_c}{2\pi}}}
        $
        &
        $\displaystyle
        h_{1L}^\perp \sim \mathcal{O}(x^0)
        $
        \\

        \addlinespace[3pt]

        \rowcolor{gluonrow}
        $\boldsymbol{T}$
        &
        $\displaystyle
        f_{1T}^\perp \sim \frac{1}{x}
        $
        &
        $\displaystyle
        g_{1T} \sim \mathcal{O}(x^0)
        $
        &
        $\displaystyle
        h_1 \sim h_{1T}^\perp \sim \frac{1}{x}
        $
        \\

        \bottomrule
    \end{tabular}
    \caption{Known small-$x$ asymptotics of the gluon TMDs. The first factor gives the eikonal order at which the corresponding TMD appears, while the second factor is a consequence of high-energy evolution. The $T$-odd entries $f_{1T}^\perp$, $h_1$ and $h_{1T}^\perp$ hold for the dipole-type distributions, given by the odderon, while the WW-type distributions have no eikonal contribution (Sec.~\ref{subsub:sivers}). The $g_{1L}$ entry holds for both gauge-link structures in the linearized double-logarithmic approximation at large $N_c$~\cite{Kovchegov:2025gcg}. The WW-type $h_{1L}^\perp$ vanishes at eikonal order and possibly at sub-eikonal order~\cite{Kovchegov:2025gcg}, and $g_{1T}$ has not been computed at small $x$, so that these two entries reflect naive power counting.}
    \label{tab:gluon-TMD-smallx}
\end{table}

The quark sector is better understood, since small-$x$ operator definitions are available for all eight leading-twist quark TMDs, each at the first eikonal order at which it does not vanish~\cite{Kovchegov:2022kyy,Santiago:2023rfl,Adamiak:2024khm}. These build on the polarized Wilson-line operator constructed to sub-sub-eikonal accuracy in the study of the quark Sivers function~\cite{Kovchegov:2021iyc}, which was extended to the Boer--Mulders function~\cite{Kovchegov:2022kyy}. These orders differ: the Sivers function receives an eikonal contribution from the odderon (see Sec.~\ref{subsub:sivers}), whereas the helicity distributions, like their gluon counterparts, first appear at sub-eikonal order. In general, it is also useful to distinguish the flavor-singlet case, in which the unpolarized quark TMD is eikonal (see Sec.~\ref{subsub:inclusive}), from the flavor non-singlet case, in which it is suppressed by a power of $x$. In both cases, the quark helicity TMDs have been computed within the CGC~\cite{Kovchegov:2016zex}.

\subsection{High-energy evolution of spin-dependent subeikonal TMDs}
\label{sub:smallx-helicity}

Once such tree-level expressions have been obtained, an important question is the $x$ dependence of the operators generated by high-energy evolution. Addressing this requires calculating the one-loop corrections to polarized Wilson lines in the regime where the emitted gluon carries a small longitudinal momentum fraction. Two main features of this high-energy evolution should be highlighted: (i) unlike the evolution of eikonal Wilson lines, which is single logarithmic and resums powers of $\alpha_s\ln(1/x)$, the evolution of polarized Wilson lines relevant for helicity and, more generally, subeikonal polarized TMDs is double logarithmic at leading logarithmic accuracy, resumming powers of $\alpha_s\ln^2(1/x)$. (ii) As in the Balitsky hierarchy, the evolution does not close on a finite set of operators in general. In addition, the evolution mixes polarized Wilson-line operators of different types, such as the gluonic and quark operators that enter the helicity sector. 

In the large-$N_c$ limit, or in the combined large-$N_c$ and large-$N_f$ limit, the system of evolution equations closes and can therefore be solved either numerically or analytically. Furthermore, to determine the leading small-$x$ asymptotic behaviour, it is sufficient to work at double-logarithmic accuracy, without resumming $\alpha_s\ln(1/x)$ corrections to all orders. The resulting small-$x$ asymptotic behaviour of all gluon and quark TMDs known to date is summarized in Tables~\ref{tab:gluon-TMD-smallx}--\ref{tab:quark-FNS-TMD-smallx}. In these tables, the eikonal order follows from setting $\alpha_s=0$, which amounts to neglecting the high-energy evolution. A particularly notable result is the small-$x$ behaviour of the helicity TMDs in both the quark and gluon sector:
\begin{align}
    g_1(x,\kt)\sim x^{-3.66\sqrt{\frac{\alpha_sN_c}{2\pi}}}C_{\textrm{sub-eik}}(\kt) \,.
\end{align}
This result has been obtained numerically~\cite{Cougoulic:2022gbk} and analytically~\cite{Borden:2023ugd}, in close agreement with the earlier infrared evolution equation result~\cite{Bartels:1996wc}, the two intercepts differing in the third decimal place. The DGLAP/BFKL duality provides a further cross-check in the large-$N_c$ and large-$N_f$ limit, where the intercept becomes $3.48\sqrt{\alpha_sN_c/2\pi}$ for $N_f=N_c=3$~\cite{Borden:2025ehe}. The same intercept governs the WW-type gluon helicity distribution, whose evolution equation coincides with that of the dipole-type distribution in the linearized double-logarithmic approximation at large $N_c$~\cite{Kovchegov:2025gcg}. The two are expected to differ in the saturation region, where the single-logarithmic evolution of the WW operator does not close on the polarized dipole amplitudes. Note that the quark and gluon orbital angular momentum distributions share the same intercept (Sec.~\ref{sec:gtmd}).

In addition, the small-$x$ asymptotics of all eight leading-twist quark TMDs are now known at large $N_c$ in the linearized double-logarithmic approximation, for the flavor non-singlet~\cite{Santiago:2023rfl} and the flavor singlet~\cite{Adamiak:2024khm} cases, and are collected in Tables~\ref{tab:quark-FS-TMD-smallx} and~\ref{tab:quark-FNS-TMD-smallx}. The flavor singlet sector dominates for the $T$-even distributions, whereas the Sivers function is dominated by the flavor non-singlet sector through the eikonal odderon, and the chiral-odd distributions have identical asymptotics in the two sectors since they do not mix with gluons~\cite{Adamiak:2024khm}. For the helicity distributions, the polarized DGLAP anomalous dimensions resummed by small-$x$ evolution have been extracted, and they agree with the existing fixed-order calculations up to three loops~\cite{Borden:2023ugd,Borden:2024bxa,Borden:2025ehe}.

All these results neglect quark-mass terms and saturation, and their matching onto CSS evolution has not yet been studied. Understanding Sudakov effects for these subeikonal, spin-dependent TMDs within the CGC framework is an important direction for future research. 

Small-$x$ helicity evolution has been used in global extractions of the helicity distributions, where it fixes the $x$ dependence below the reach of the data instead of extrapolating a parametrization. A first analysis of polarized DIS data at $x<0.1$ reduced the small-$x$ uncertainty on $g_1^p$ to about $25$ percent, against about $100$ percent for a DGLAP-based extrapolation~\cite{Adamiak:2021ppq}. Including SIDIS data with the revised evolution gave $\int_{10^{-5}}^{0.1}\der x\,(\tfrac12\Delta\Sigma+\Delta G)=-0.64\pm0.60$ at $Q^2=10$~GeV$^2$~\cite{Adamiak:2023yhz}, and adding the polarized jet data from the Relativistic Heavy-Ion Collider (RHIC) reduced the corresponding integral over $10^{-7}<x<0.1$ to $-0.04\pm0.23$~\cite{JAMCollaborationSmall-xAnalysisGroup:2025tfa}. Initial conditions taken from a valence-quark model of the proton~\cite{Dumitru:2024pcv} give instead positive values, between $0.6$ and $1.4$ depending on the running-coupling prescription~\cite{Adamiak:2025mdy}. 

\begin{table}[t]
   \centering
    \renewcommand{\arraystretch}{1.8}
    \setlength{\tabcolsep}{7pt}

    \begin{tabular}{
        >{\centering\arraybackslash}m{2.0cm}
        >{\centering\arraybackslash}m{3.7cm}
        >{\centering\arraybackslash}m{3.7cm}
        >{\centering\arraybackslash}m{3.7cm}
    }
        \rowcolor{tmdteal}
        \multicolumn{4}{c}{
            \color{white}\large\bfseries
            Flavor-singlet quark TMD distributions
        } \\

        \rowcolor{tmdlight}
        \multicolumn{1}{c}{\textbf{Nucleon}}
        &
        \multicolumn{3}{c}{\textbf{Quark polarization}} \\

        \rowcolor{tmdlight}
        \textbf{polarization}
        & \textbf{Unpolarized}
        & \textbf{Longitudinal ($L$)}
        & \textbf{Transverse ($T$)} \\

        \midrule

        \rowcolor{tmdrow}
        $\boldsymbol{U}$
        &
        $\displaystyle
        f_1 \sim \frac{1}{x}
        x^{-\frac{4\alpha_s N_c\ln 2}{\pi}}
        $
        &
        \textemdash
        &
        $\displaystyle
        h_1^\perp \sim x
        $
        \\

        \addlinespace[3pt]

        $\boldsymbol{L}$
        &
        \textemdash
        &
        $\displaystyle
        g_1 \sim
        x^0 x^{-3.66\sqrt{\frac{\alpha_s N_c}{2\pi}}}
        $
        &
        $\displaystyle
        h_{1L}^\perp \sim x
        $
        \\

        \addlinespace[3pt]

        \rowcolor{tmdrow}
        $\boldsymbol{T}$
        &
        $\displaystyle
        f_{1T}^\perp \sim
        x^0 x^{-2.9\sqrt{\frac{\alpha_s N_c}{4\pi}}}
        $
        &
        $\displaystyle
        g_{1T} \sim
        x^0 x^{-2.9\sqrt{\frac{\alpha_s N_c}{4\pi}}}
        $
        &
        $\displaystyle
        h_1 \sim h_{1T}^\perp \sim
        x x^{-2\sqrt{\frac{\alpha_s N_c}{2\pi}}}
        $
        \\

        \bottomrule
    \end{tabular}
    \caption{Same as table~\ref{tab:gluon-TMD-smallx} for the quark sector in the flavor singlet case, at large $N_c$ in the linearized double-logarithmic approximation~\cite{Adamiak:2024khm}.}
    \label{tab:quark-FS-TMD-smallx}
\end{table}

\begin{table}[t]   
    \centering
    \renewcommand{\arraystretch}{1.8}
    \setlength{\tabcolsep}{7pt}

    \begin{tabular}{
        >{\centering\arraybackslash}m{2.0cm}
        >{\centering\arraybackslash}m{3.7cm}
        >{\centering\arraybackslash}m{3.7cm}
        >{\centering\arraybackslash}m{3.7cm}
    }
        \rowcolor{nsgold}
        \multicolumn{4}{c}{
            \color{white}\large\bfseries
            Flavor non-singlet quark TMD distributions
        } \\

        \rowcolor{nsgoldlight}
        \multicolumn{1}{c}{\textbf{Nucleon}}
        &
        \multicolumn{3}{c}{\textbf{Quark polarization}} \\

        \rowcolor{nsgoldlight}
        \textbf{polarization}
        & \textbf{Unpolarized}
        & \textbf{Longitudinal ($L$)}
        & \textbf{Transverse ($T$)} \\

        \midrule

        \rowcolor{nsgoldrow}
        $\boldsymbol{U}$
        &
        $\displaystyle
        f_1 \sim
        x^0 x^{-\sqrt{\frac{2\alpha_s C_F}{\pi}}}
        $
        &
        \textemdash
        &
        $\displaystyle
        h_1^\perp \sim x
        $
        \\

        \addlinespace[3pt]

        $\boldsymbol{L}$
        &
        \textemdash
        &
        $\displaystyle
        g_1 \sim
        x^0 x^{-\sqrt{\frac{\alpha_s N_c}{\pi}}}
        $
        &
        $\displaystyle
        h_{1L}^\perp \sim x
        $
        \\

        \addlinespace[3pt]

        \rowcolor{nsgoldrow}
        $\boldsymbol{T}$
        &
        $\displaystyle
        f_{1T}^\perp \sim \frac{1}{x}
        $
        &
        $\displaystyle
        g_{1T} \sim x^0
        $
        &
        $\displaystyle
        h_1 \sim h_{1T}^\perp \sim
        x x^{-2\sqrt{\frac{\alpha_s N_c}{2\pi}}}
        $
        \\

        \bottomrule
    \end{tabular}
    \caption{Same as table~\ref{tab:gluon-TMD-smallx} for the quark sector in the flavor non-singlet case, in the linearized double-logarithmic approximation~\cite{Santiago:2023rfl}. For the quark Sivers function the effect of quantum evolution to the sub-eikonal correction is also known~\cite{Kovchegov:2022kyy}.}
    \label{tab:quark-FNS-TMD-smallx}
\end{table}

\subsection{Subeikonal corrections to eikonal TMDs}
\label{sec:subeik-eik}

The progress of the CGC effective field theory in accounting for subeikonal corrections has also benefited TMDs that are not power suppressed in the small-$x$ limit, particularly in the technically simpler quark sector. This is the case for the flavor non-singlet quark Sivers function. The developments discussed in the previous subsection have enabled one to obtain the small-$x$ asymptotic behaviour of the subeikonal correction to the flavor non-singlet Sivers function, such that
\begin{align}
    f_{1T}^\perp(x,\kt)\sim \frac{1}{x}C_{\rm eik}(\kt)+x^{-3.4\sqrt{\frac{\alpha_sN_c}{4\pi}}}C_{\textrm{sub-eik}}(\kt)+... \,.
\end{align}
The first term is the odderon contribution which we discussed in section~\ref{subsub:sivers} while the second term associated with the subeikonal correction to the Sivers function has been computed in~\cite{Kovchegov:2022kyy}.

For spin-independent TMDs, the subeikonal operator definition was obtained in~\cite{Altinoluk:2023qfr,Altinoluk:2024tyx}, considering back-to-back dijet production in DIS and forward proton--nucleus scattering, respectively. As for spin-dependent TMDs, the operator definition at subeikonal accuracy involves an insertion of the quark background field. Several questions remain open. In particular, it is necessary to understand how to model, within the CGC framework, the correlator of such quark background fields, as well as how the corresponding operator evolves at high energy. Next-to-eikonal corrections from $t$-channel quark exchange to DIS and SIDIS have been expressed through the collinear quark distributions and the unpolarized quark TMD~\cite{Altinoluk:2025ang}. A Gaussian model for the corresponding decorated dipole and quadrupole operators has been proposed~\cite{Agostini:2025vvx}, and the rapidity evolution of the operator with a background-quark insertion has been shown to reproduce quark reggeization~\cite{Altinoluk:2026ymh,Altinoluk:2026ngv}. In the gluon sector, a closed non-linear evolution equation for the sub-eikonal dipole gluon distribution has been derived at large $N_c$~\cite{Li:2026azt}.

A complementary approach starts from moderate $x$. In the background-field method, the quark propagator in the target field is expanded in transverse gradients while the full longitudinal phase $e^{-ixP^-z^+}$ is retained~\cite{Mukherjee:2023snp,Kar:2026vzk}. Applied to back-to-back dijet production in DIS, this expansion organizes the leading-order cross section at arbitrary $x$ in terms of gluon TMD operators to twist-three accuracy, including three-gluon correlators and insertions of $F^{+-}$ and $F^{ij}$, and in the small-$x$ limit it reproduces the sub-eikonal CGC results for the same process~\cite{Mukherjee:2026cte,Altinoluk:2022jkk,Altinoluk:2024zom}. The equations of motion then reduce the number of independent twist-three matrix elements. The contributions of the quark background, which enter through quark--gluon and quark--antiquark dijets, have been obtained to the same accuracy~\cite{Mukherjee:2026six}. At small $x$, the leading-twist quark operator coincides with the one identified at sub-eikonal accuracy in the CGC~\cite{Altinoluk:2023qfr}, whereas the twist-three quark contributions correspond to sub-sub-eikonal corrections that have not yet been computed in the CGC. An operator basis that contains both the all-twist eikonal limit and the leading-twist limit at arbitrary $x$ has also been constructed for unpolarized dijet production in DIS~\cite{Kar:2026cnx}. These results are at leading order in $\alpha_s$ and do not include the high-energy evolution of the operators, so that their connection with saturation dynamics beyond the classical field approximation remains to be established. 

\section{Observables and phenomenology}
\label{sec:pheno}

In this section we review the observables through which small-$x$ TMDs are probed, the comparisons with data available so far, and the signatures of saturation expected at RHIC, the Large Hadron Collider (LHC) and the EIC. We proceed from semi-inclusive DIS and two-particle correlations to heavy flavor, diffraction and energy correlators.

\subsection{Semi-inclusive DIS and lepton--jet correlations}
\label{sec:sidis-pheno}

The simplest observable sensitive to small-$x$ distributions is SIDIS in the regime where the transverse momentum of the measured hadron is much smaller than the photon virtuality, $k_\perp\ll Q$. In that limit the cross section factorizes at leading power onto a hard factor and a sea-quark distribution \cite{Marquet:2009ca,Xiao:2010sa}. The expected signal is a suppression of the nuclear modification factor, the ratio of the $eA$ to the $ep$ cross section normalized by the mass number, at hadron transverse momenta of the order of the saturation scale. A computation within the CGC, validated against HERA data, predicts a significant suppression at the EIC for $x\lesssim10^{-3}$, which becomes stronger as $x$ decreases and the saturation scale grows, and depends only weakly on $Q^2$~\cite{Casuga:2026dif}. Since the hadron transverse momentum and rapidity can be varied in addition to $x$, $Q^2$ and the nuclear mass number, SIDIS at the EIC can follow the transition from the dilute to the saturated regime and constrain the initial condition of the BK evolution. These predictions are at leading order in $\alpha_s$~\cite{Casuga:2026dif}, and we expect them to be modified by higher-order corrections, in particular by the resummation of Sudakov logarithms.

Lepton--jet correlations in the lab frame access the same distribution with a different experimental handle. At leading order the lepton and the jet are back-to-back, and their imbalance $\boldsymbol{q}_\perp$ is the transverse momentum of the struck quark, smeared by soft radiation from the jet, so that the cross section factorizes onto the quark TMD~\cite{Liu:2018trl,Liu:2020dct}, which at small $x$ is the sea-quark distribution of Sec.~\ref{subsub:inclusive}. In the CGC, the azimuthal harmonics of the correlation are suppressed at small imbalance, more strongly for a gold than for a proton target and in contrast with collinear-framework predictions, which makes them a probe of saturation~\cite{Tong:2022zwp}. The measurement has been performed at HERA~\cite{H1:2021wkz}, but at $Q^2>150$~GeV$^2$ it probes $x$ above a few $10^{-3}$ and is described by moderate-$x$ TMD and fixed-order calculations, so that it tests the framework and not saturation.

Sea-quark distributions also enter two-particle correlations. When the projectile parton scatters off a small-$x$ sea quark, as in quark--gluon dijets and photon plus quark jet production in DIS, or photon plus gluon jet and quark--gluon dijet production in forward $pA$ collisions, the leading-order CGC calculation reproduces TMD factorization with the hard factors of moderate $x$~\cite{Caucal:2025xxh}. At large $N_c$ the sea-quark distributions that appear in these channels are the SIDIS distribution convoluted with one or more dipole correlators, a structure anticipated from the gauge-link analysis in~\cite{Xiao:2010sa}. Each additional dipole enlarges the effective saturation scale, and generates a Cronin-like peak in the nuclear modification that is absent for the SIDIS distribution. With the exception of photon plus jet production in DIS, where the sea-quark channel is the leading one, these channels are suppressed by one power of $\alpha_s$ relative to the gluon-initiated channels with the same final state. They share the energy dependence of those channels, however, and have not yet been included in the phenomenology of forward dihadrons or photon--jet pairs in $pA$ collisions. 

\subsection{Two-particle correlations and the gluon distributions}
\label{sec:twoparticle}

The color flow of the partonic subprocess determines the gauge link structure of the participating TMD. In forward dihadron and dijet production in $pp$ and $pA$ collisions, colored partons are present in both the initial and the final state. Several gluon TMDs with different gauge-link structures contribute \cite{Dominguez:2011wm,Kotko:2015ura}. In the large-$N_c$ limit these reduce to convolutions of the dipole and WW distributions, so that the proliferation of operators does not bring with it a proliferation of independent non-perturbative functions~\cite{Dominguez:2012ad,Marquet:2016cgx}.

The phenomenology is most developed for forward dihadrons in $pA$ collisions. Forward production selects a large-$x$ parton from the proton and a small-$x$ parton from the nucleus, which is the kinematics of the hybrid dilute-dense formalism~\cite{Dumitru:2005gt,Altinoluk:2011qy,Chirilli:2011km}. In the back-to-back limit the imbalance of the pair is set by the transverse momentum of the gluon TMD, and is therefore sensitive to saturation. The measured quantity is the coincidence probability, the number of pairs per trigger as a function of the relative azimuthal angle. Saturation acts on it in two ways. Multiple scattering in the classical field broadens the away-side peak, and non-linear evolution suppresses it~\cite{Kharzeev:2003wz,Albacete:2003iq}. Both grow with the saturation scale, and therefore with energy, with forward rapidity, with nuclear size and with centrality~\cite{Kharzeev:2004bw,Marquet:2007vb,Albacete:2010pg}. CGC calculations with a Gaussian distribution of color sources predict both effects, first with the quark-initiated channel alone~\cite{Lappi:2012nh} and later with the gluon channel included~\cite{Albacete:2018ruq}.

The suppression is seen and the broadening is not. STAR and PHENIX measured the suppression at RHIC in forward $d$+Au relative to $pp$~\cite{Braidot:2010ig,Adare:2011sc,STAR:2021fgw}, but the widths are comparable in the two systems and larger than leading-order CGC evolution predicts, and forward dijets at the LHC show the same pattern~\cite{ATLAS:2019jgo}. Descriptions outside the CGC account for the data as well, among them an approach based on collinear factorization and nuclear parton distributions~\cite{Perepelitsa:2025qpz}. The likely explanation for the missing broadening is that the width is set by parton showering and by the intrinsic transverse momentum from fragmentation rather than by the target, and a study including both found it insensitive to the intrinsic transverse momentum of the small-$x$ gluon~\cite{Cassar:2025vdp}.

The small-$x$ TMD framework accommodates the competing mechanisms. The back-to-back kinematics enhances the Sudakov double and single logarithms of Sec.~\ref{sec:sudakov} alongside the logarithms of $1/x$, and the joint resummation of the two towers has been proposed~\cite{Mueller:2012uf,Mueller:2013wwa,Zhou:2018lfq}. The two accounts of the missing broadening overlap to a large extent: the perturbative Sudakov logarithms resum much of the radiation generated by a parton shower~\cite{Hautmann:2017xtx,Hautmann:2017fcj}, and the non-perturbative Sudakov factor plays a role similar to the intrinsic transverse momentum from fragmentation. A first implementation, using a saturation model with parametrized $x$ and $A$ dependence, reproduced the absence of broadening, although self-normalizing the correlation discarded the information on suppression~\cite{Stasto:2018rci}.

\begin{figure}[t]
\centering
\includegraphics[width=0.75\textwidth]{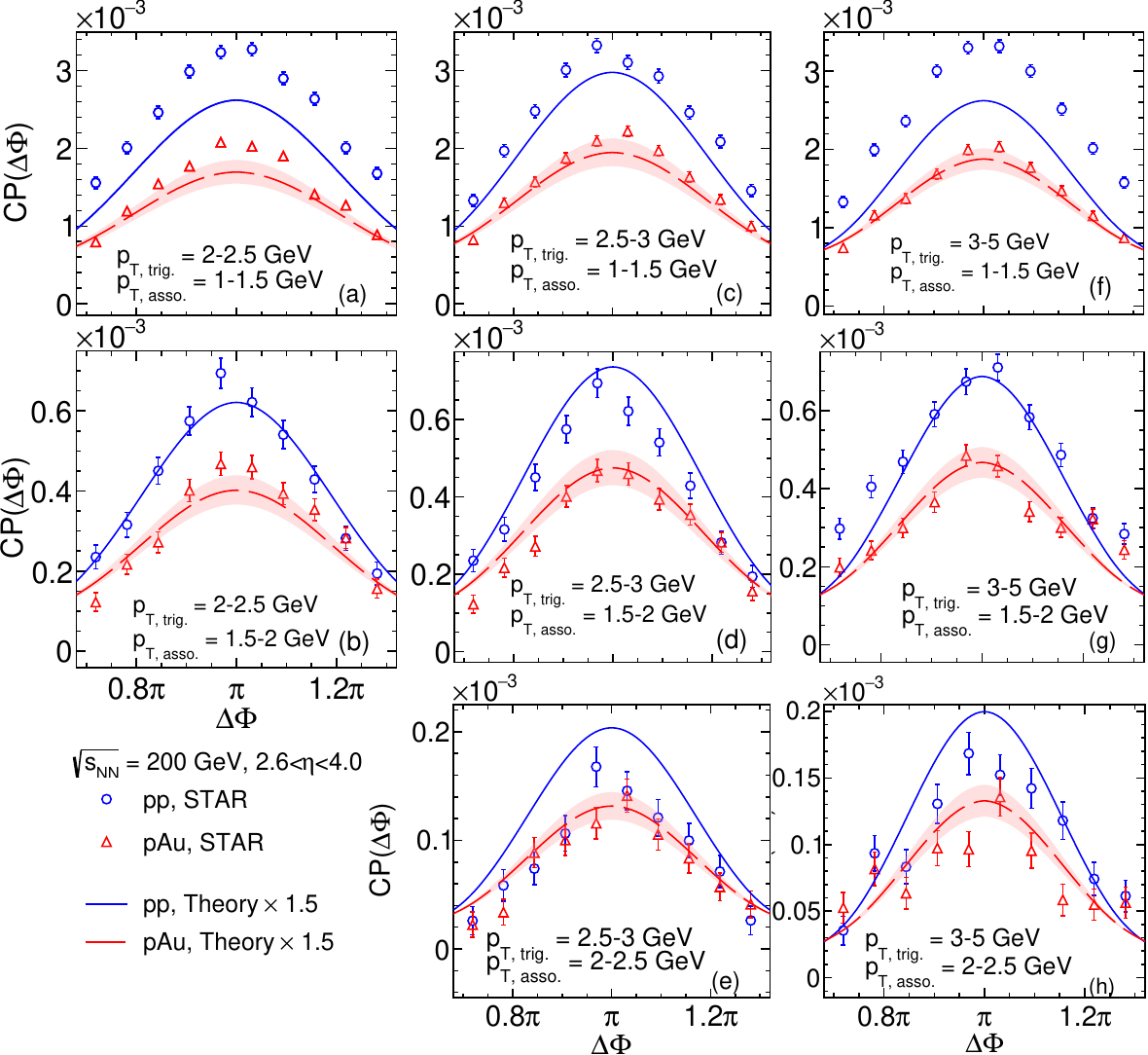}
\caption{Away-side coincidence probability of forward di-$\pi^0$ production ($2.6<\eta<4.0$) in $pp$ (blue) and $p$Au (red) collisions at $\sqrt{s_{NN}}=200$~GeV, for several bins of trigger and associated transverse momenta, compared with STAR data~\cite{STAR:2021fgw}. The calculation combines running-coupling BK evolution, perturbative and non-perturbative Sudakov resummation and ITMD hard factors, and the theory curves are multiplied by an overall factor of $1.5$. The bands correspond to a variation of $Q^2_{s0,{\rm Au}}$ by a factor of $1.2$. Figure from~\cite{Caucal:2025zkl}.}
\label{fig:STAR}
\end{figure}

The most complete calculation to date combines running-coupling BK evolution, perturbative and non-perturbative Sudakov resummation, improved TMD (ITMD) hard factors (see Sec.~\ref{sec:itmd}), and initial conditions constrained by HERA data for the proton and by a nuclear dipole model with transverse density for the nucleus, and confronts the STAR $pp$ and $p$+Au measurements~\cite{Caucal:2025zkl}, as shown in Fig.~\ref{fig:STAR}. Fragmentation controls the coincidence probability itself, whereas the ratios of the integrated area and of the width between collision systems are robust against the choice of fragmentation function and are the more reliable probes of the initial state. The predictions for the acceptance of the ALICE Forward Calorimeter (FoCal)~\cite{ALICE:2019pjv} show less suppression than at RHIC in spite of the larger saturation scale, because the smaller average fragmentation fraction shifts the partonic transverse momenta toward the dilute region.

In DIS the virtual photon is colorless, and inclusive back-to-back dijet production probes the WW distribution~\cite{Dominguez:2010xd,Dominguez:2011wm}. Leading-order studies assessed the EIC sensitivity of dihadron correlations, with an away-side suppression in $e$Au relative to $ep$~\cite{Zheng:2014vka}, and of the $\cos2\phi$ modulation induced by the linearly polarized WW distribution~\cite{Dumitru:2015gaa,Dumitru:2018kuw}. The correlation limit fails when the dijet momentum is not large compared with the imbalance~\cite{Mantysaari:2019hkq}. The process is now known at NLO~\cite{Caucal:2021ent,Taels:2022tza,Caucal:2022ulg,Bergabo:2022tcu,Bergabo:2023wed}, and its back-to-back limit factorizes into the WW distribution, a soft factor and a coefficient function~\cite{Caucal:2023nci}. In EIC kinematics the Sudakov suppression dominates the imbalance distribution below about $1.5$~GeV and washes out the Cronin peak of the leading-order nuclear modification factor, which small-$x$ evolution suppresses further, while the $\cos2\phi$ modulation stays below two percent and changes sign relative to leading order~\cite{Caucal:2023fsf}. Dihadrons avoid the jet definition but require TMD fragmentation functions for the soft factor to remain universal~\cite{Caucal:2024nsb}, and semi-inclusive single jets provide a complementary probe~\cite{Caucal:2024cdq}. 

Forward photon--jet production in $pp$ and $pA$ collisions probes the dipole distribution alone at leading order, since the photon carries no color~\cite{Dominguez:2010xd,Dominguez:2011wm}, and in $pA$ collisions the bremsstrahlung component dominates over direct photons~\cite{Gelis:2002ki}. The photon is measured directly, so its momentum is not smeared by fragmentation as a hadron trigger would be, although a fragmentation-photon component remains and has to be suppressed by isolation. CGC calculations predict a suppression of the away-side peak in forward photon--hadron~\cite{Jalilian-Marian:2012wwi} and photon--jet production, the latter of about $40$ percent for jet momenta above $5$~GeV in the FoCal acceptance~\cite{Ganguli:2023joy}. At midrapidity, the isolated photon--hadron correlations measured in $pp$ and $p$Pb collisions are described only once Sudakov resummation is included~\cite{Benic:2022ixp}. Since only the dipole distribution enters, the ITMD and HEF formulas coincide for this process~\cite{Altinoluk:2021ygv}, and the $\cos2\phi$ modulation of virtual-photon--jet pairs generated by the maximally polarized dipole gluons is strongly reduced by Sudakov radiation~\cite{Boer:2017xpy}. A complete NLO calculation of forward photon--jet production in $pA$ collisions is not yet available. In DIS, photon--jet correlations show a suppression and broadening of the back-to-back peak from proton to gold targets~\cite{Kolbe:2020tlq}, and in forward $pA$ collisions the same FoCal acceptance gives access to dijets, whose nuclear suppression is not altered by the Sudakov factor~\cite{Al-Mashad:2022zbq}.

Although, strictly speaking, this is not a two-particle correlation, we note that trijet photoproduction is sensitive to the WW gluon TMD~\cite{Altinoluk:2020qet}, while dijet plus photon production in forward $pA$ collisions probes gluon TMDs with more complicated gauge-link structures~\cite{Altinoluk:2018byz}. In both cases, TMD factorization at leading order has been derived from the CGC.

\subsection{Heavy flavor}
\label{sec:heavyflavor}

Heavy-quark pairs differ from light dijets in their access to linearly polarized gluons. In forward $gA\to Q\bar Q$ production and in photoproduction $\gamma A\to Q\bar Q$, the leading-order coefficient of the linearly polarized distribution in the $\cos2\phi$ modulation is proportional to the squared quark mass, so it vanishes for light dijets and dihadrons and survives for charm and bottom pairs~\cite{Akcakaya:2012si,Marquet:2017xwy}. Light dijets in DIS retain this sensitivity only through the photon virtuality, and the corresponding heavy-quark and dijet asymmetries at the EIC were studied in~\cite{Boer:2010zf,Boer:2016fqd}. In forward $pA$ collisions three unpolarized and three linearly polarized gluon distributions contribute, and the adjoint dipole among them is maximally polarized~\cite{Marquet:2017xwy}. 

The mass also supplies a perturbative scale that, for charm, is comparable to the saturation scale, and it changes the structure of the Sudakov resummation. For imbalances below the quark mass, the dead cone cuts off collinear final-state radiation, so that the final-state Sudakov factor becomes single logarithmic, while the light-quark double logarithm is recovered when the mass is small compared with the imbalance~\cite{Marquet:2025jdr}. This unified resummation describes the H1 data on $D^*$--jet and $D^*$--dijet azimuthal correlations, and predicts for ultra-peripheral collisions at the LHC an away-side nuclear ratio of about $0.85$ for $D$ pairs and $0.80$ for $B$ pairs. Its extension to forward $pA$ collisions describes the LHCb $D^0\bar D^0$ correlations in $pp$ and $p$Pb collisions, with a suppression that grows with rapidity and is stronger for bottom than for charm~\cite{Gao:2026azd,LHCb:2020jse}. In the resummed calculations available, Sudakov radiation suppresses the linearly polarized contribution, and the $\cos2\phi$ modulation is then dominated by soft radiation acting on the unpolarized distribution~\cite{Boer:2017xpy,Hatta:2021jcd,Marquet:2025jdr,Gao:2026azd}. The access to linear polarization is tied to the hard factor, and the handle on saturation lies in ratios between targets, between harmonics and between quark masses rather than in the size of a single asymmetry. Photonuclear $D^0$ production in ultra-peripheral collisions has been measured by CMS~\cite{CMS:2025jjx}.

Direct quarkonium production in DIS has been formulated in a joint CGC and non-relativistic QCD (NRQCD) framework, both with an associated gluon~\cite{Kang:2023doo} and for the full set of NRQCD channels~\cite{Cheung:2024qvw}. In the correlation limit, the color-octet channels factorize onto the WW distribution, in agreement with the TMD calculation~\cite{Bacchetta:2018ivt}, and the linearly polarized WW distribution appears in the azimuthal correlation between the scattered electron and the quarkonium~\cite{Bacchetta:2018ivt,DAlesio:2019qpk,Cheung:2024qvw}. Genuine saturation corrections beyond the TMD limit suppress the yield at low transverse momentum when the saturation scale approaches the hard scale~\cite{Cheung:2024qvw}. At the LHC, quarkonium-pair production has been proposed as a probe of linearly polarized gluons in the proton, so far within moderate-$x$ TMD factorization~\cite{Lansberg:2017dzg,Scarpa:2019fol}.

\subsection{Diffractive observables}
\label{sec:diffpheno}

The factorization of diffractive processes and the gluon diffractive distribution were discussed in Sec.~\ref{sec:tmds}. Their observable consequence is that the imbalance of a hard diffractive dijet produced through the (2+1)-jet mechanism lies near the saturation scale of the Pomeron~\cite{Iancu:2021rup,Iancu:2022lcw}, while the semi-hard third jet emerges two to three units of rapidity away, toward the target~\cite{Iancu:2023lel}. In ultra-peripheral collisions at the LHC this channel scales as $A$, whereas exclusive dijets scale as $A^{4/3}$ but are suppressed by two to three orders of magnitude, and requiring dijet momenta above $20$~GeV restricts the Pomeron to $x_{\mathbb{P}}\sim10^{-2}$, which is only marginal for saturation~\cite{Iancu:2023lel}. The quark channel enters diffractive SIDIS and diffractive dijets with a soft quark, which share the same quark diffractive distribution~\cite{Hatta:2022lzj,Hauksson:2024bvv}. In incoherent diffraction, where the target breaks up, the relevant distributions saturate logarithmically for momentum transfers below the saturation scale and fall as $1/|t|^2$ above it~\cite{Rodriguez-Aguilar:2023ihz,Rodriguez-Aguilar:2024efj}. With heavy quarks, the quark diffractive distribution of a nucleus develops a Cronin peak near $1$~GeV that the charm mass strongly suppresses, so that comparing charm- and bottom-tagged diffractive jets offers a handle on saturation, although CSS evolution is expected to modify these predictions significantly~\cite{Caucal:2026xhj}.

Radiation is the main obstacle in this channel. CSS evolution moves the gluon diffractive distribution away from the saturation plateau, so that its peak lies at $1$ to $2$~GeV, well below the imbalances of $6$~GeV and above probed by the CMS measurement~\cite{Iancu:2025jsu}. Including DGLAP evolution in the boundary condition makes the distribution negative at large momentum. Initial-state radiation broadens the imbalance and reduces the $\cos2\phi$ modulation~\cite{Shao:2024nor}, and the dijet acoplanarity and the transverse energy--energy correlator have been proposed as more robust variables~\cite{Shao:2026doo}. CMS measured a positive $\cos2\Phi$ that rises with the imbalance in photoproduced dijets with a rapidity gap in Pb+Pb collisions~\cite{CMS:2022lbi}, which soft-gluon radiation reproduces on average~\cite{Hatta:2020bgy}, and the comparison with the resummed diffractive calculation still shows a sizable discrepancy~\cite{Shao:2024nor}. ATLAS has measured the cross section for photon--Pomeron collisions into jets~\cite{ATLAS:2026cxr}, so far compared only with collinear NLO calculations~\cite{Guzey:2016tek}.

Exclusive dijets also probe the spatial structure of the target. The correlation between the dijet momentum and the momentum transfer, measured by $\cos2(\phi_P-\phi_\Delta)$, is sensitive to the elliptic gluon Wigner distribution~\cite{Altinoluk:2015dpi,Hatta:2016dxp,Hagiwara:2017fye}. With a JIMWLK-evolved proton the modulation is at the percent level at the EIC and decreases with decreasing $x$~\cite{Mantysaari:2019csc}, each harmonic of the cross section is fixed by the corresponding harmonic of the dipole~\cite{Salazar:2019ncp}, and in incoherent diffraction the modulation can reach tens of percent~\cite{Mantysaari:2019hkq}. In the (2+1)-jet channel the dijet momentum and the imbalance are uncorrelated at leading order~\cite{Hauksson:2024bvv}, and the diffractive linearly polarized gluon distribution vanishes~\cite{Hatta:2024vzv}, so that imaging requires the momentum transfer to the target to be measured, as foreseen with the far-forward detectors of the EIC.

\subsection{New observables: energy correlators and hadronic recoil}
\label{sec:recoil}

Energy correlators weight the final state by the energy deposited at a given angle. They are infrared and collinear safe, and at leading power the fragmentation functions drop out through the momentum sum rule, so that they probe the flow of energy rather than identified hadrons. At small $x$, the nucleon energy correlator is suppressed at small angles relative to collinear factorization, and the angle at which it turns over reflects the saturation scale~\cite{Liu:2023aqb}. The one-point correlator measured as a function of the angle to the target depends on the dipole as its only non-perturbative input and shows a strong nuclear suppression at small angles~\cite{Kang:2026hig}. Back-to-back transverse energy--energy correlators provide an infrared-safe version of the imbalance, for electron--hadron pairs at the EIC, where the nuclear modification factor is suppressed at small angles~\cite{Kang:2023oqj}, and for photon--hadron pairs in $pp$ and $p$Pb collisions, where the suppression survives at hadron level~\cite{Kang:2025vjk,Ganguli:2025aqa}.

While these suppressions are sizable, comparable ones are obtained with collinear nuclear distributions supplemented by transverse-momentum broadening~\cite{Kang:2024otf,Fu:2025hpc}, and the modification of in-jet energy correlators in $p$Pb collisions is described without saturation~\cite{Barata:2024wsu,ALICE:2026giw}. The size of a suppression therefore does not establish saturation by itself, and its dependence on angle, on $x$ and on nuclear size is required.

Showering and hadronization also redistribute energy in angle. A recent event-generator study suggests that the parton-level nuclear broadening can be strongly diluted in the nucleon energy correlator and in the dihadron decorrelation, whereas the global hadronic recoil appears to retain it~\cite{Ke:2026ytl}, and a similar washing out affects the quark fracture function~\cite{Caucal:2026zpt}. Recoil observables that subtract the radiation inside a rapidity interval reduce the sensitivity to soft radiation~\cite{Lin:2026typ,Li:2026uug}. These observables are known at leading order or leading-logarithmic accuracy, and their small-$x$ NLO corrections remain to be computed.

\section{Related topics and interdisciplinary connections}

\label{sec:advanced}

In this last section, we discuss several topics that connect small-$x$ TMDs to neighboring areas: generalized TMDs and orbital angular momentum, power corrections beyond leading power in TMD factorization, Monte Carlo event generators and heavy-ion collisions.

\subsection{Generalized TMDs at small $x$ and related topics}
\label{sec:gtmd}

The first subject closely connected to TMD distributions at small $x$ is their off-forward generalization, namely generalized transverse-momentum-dependent distributions (GTMDs). At vanishing skewness, the Fourier transform of a GTMD with respect to the transverse momentum transfer can be interpreted as a Wigner distribution, or quantum phase-space distribution, encoding simultaneously the longitudinal momentum fraction, transverse momentum, and transverse position of partons. At small $x$, the WW gluon GTMD reads
\begin{equation}
xG_{\rm WW}(x,\boldsymbol{k},\boldsymbol{\Delta})
=\frac{-2}{\alpha_s}
\int \frac{\der^2\boldsymbol{x}\der^2\boldsymbol{y}}{(2\pi)^4} \ 
e^{-i\boldsymbol{k}\cdot(\boldsymbol{x}-\boldsymbol{y})}
e^{i\boldsymbol{\Delta}\cdot(\boldsymbol{x}+\boldsymbol{y})/2}
\left\langle
\mathrm{Tr}
\left[
\partial_i U(\boldsymbol{x})
U^\dagger(\boldsymbol{y})
\partial_i U(\boldsymbol{y})
U^\dagger(\boldsymbol{x})
\right]
\right\rangle_x \,.
\label{eq:WW-GTMD}
\end{equation}
GTMDs give access to correlations between transverse momentum and impact parameter that are absent in TMDs and GPDs. This makes them particularly interesting at small $x$, where the high-energy dynamics can be expressed in terms of Wilson-line correlators as shown in Eq.\,\eqref{eq:WW-GTMD}. Some features of the gluon Wigner distribution, in particular its elliptic modulation, can be probed through azimuthal correlations in exclusive diffractive dijet production, where the correlation between the transverse momentum of the dijet and that of the recoiling proton reflects the phase-space structure of small-$x$ gluons~\cite{Hatta:2016dxp}.

The connection between GTMDs and orbital angular momentum (OAM) is
particularly important at small $x$. The gluon OAM distribution can
be related to the gluon Wigner distribution, connecting the transverse
phase-space structure of gluons to their orbital motion~\cite{Hatta:2016aoc}.
In the small-$x$ limit, this analysis established a relation between
the gluon OAM and helicity distributions, $L_g(x) \simeq -2\,\Delta G(x)$, which reflects the close connection between the Wilson-line operators
underlying these two distributions~\cite{Hatta:2016aoc}. This connection
was further explored by relating the small-$x$ quark and gluon OAM distributions to polarized dipole amplitudes~\cite{Kovchegov:2019rrz}. The small-$x$ behavior of the gluon OAM distribution has been studied
using different approaches to helicity evolution. In particular,
its asymptotic
behavior was obtained within the double-logarithmic approximation~\cite{Boussarie:2019icw}, with the
same leading small-$x$ intercept as for the gluon helicity distribution.
This connection was revisited in~\cite{Kovchegov:2023yzd}
using the small-$x$ helicity evolution equations discussed in section~\ref{sec:beyond-eikonal}, which yield
a common asymptotic behavior for the gluon OAM and helicity distributions,
\begin{equation}
L_G(x,Q^2)\sim \Delta G(x,Q^2)
\sim \left(\frac{1}{x}\right)^{
3.66\sqrt{\frac{\alpha_s N_c}{2\pi}}}.
\end{equation}
The two intercepts agree to two decimal places and differ slightly beyond~\cite{Borden:2023ugd,Borden:2025ehe}. At large $N_c$ and $N_f$ and $Q^2=10$~GeV$^2$, one finds $L_{q+\bar q}/\Delta\Sigma\approx-1.0$ and $L_G/\Delta G\approx-1.9$~\cite{Kovchegov:2023yzd,Manley:2024pcl,Becker:2026xcp}, close to the small-$x$ estimate $L_g\simeq-2\,\Delta G$. The small-$x$ GTMD framework also exposes nontrivial spin-orbit correlations. In particular, it has been shown that the helicity and orbital angular momentum of individual quarks and gluons are strongly anti-aligned at small $x$, even in unpolarized or spinless hadrons and nuclei~\cite{Bhattacharya:2024sno}. This provides a more detailed picture of small-$x$ partonic structure than that obtained from spin-averaged TMDs alone, and highlights the role of quantum spin-orbit correlations in the gluon-dominated regime.

The OAM connection has subsequently been developed into concrete experimental observables. In exclusive dijet production in electron-proton scattering, the longitudinal double-spin asymmetry contains a $\cos\phi$ azimuthal correlation between the scattered electron and the recoiling proton that is sensitive to the gluon OAM distribution at small $x$ and to its interplay with gluon helicity~\cite{Bhattacharya:2022vvo}. A more detailed analysis of this observable, including its sensitivity to gluon spin-orbit correlations, was presented in~\cite{Bhattacharya:2024sck}. An alternative process, semi-inclusive diffractive DIS, was proposed as a complementary probe of gluon tomography in nucleons and nuclei at small $x$~\cite{Hatta:2022lzj}, and its sensitivity to OAM and spin-orbit correlations was subsequently investigated in~\cite{Bhattacharya:2024sck}.

The small-$x$ structure of GTMDs has been studied systematically beyond the particular distributions relevant for OAM. In~\cite{Benic:2026idy}, the complete set of leading-twist gluon and sea-quark GTMDs was derived in the eikonal approximation at vanishing skewness for dipole-type gauge link configuration. A particularly striking result is that all gluon GTMDs can be expressed in terms of a single basic gluon dipole operator, including proton helicity-flip contributions, leading to nontrivial universal relations among otherwise independent GTMDs. The same framework also makes explicit the connections between GTMDs, TMDs, and GPDs in the small-$x$ limit. More generally, the shockwave formulation provides a unified description of PDFs, TMDs, GPDs, GTMDs, and diffractive distributions in terms of Wilson-line correlators within the Color Glass Condensate framework~\cite{Bhattacharya:2025fnz}.

GTMDs and their moments are also closely connected to the energy-momentum tensor (EMT) and to gravitational form factors, providing another application of small-$x$ parton tomography. In the gluon sector, the small-$x$ limit allows relations to be established between gravitational form factors and the dipole scattering amplitude. In particular, the gluon $A$-type gravitational form factor can be related to the transverse structure of the dipole amplitude~\cite{Hagiwara:2024wqz}. Gravitational transverse-momentum distributions have been introduced to study the energy-momentum tensor directly in the small-$x$ regime, giving access to gluon momentum densities as well as transverse momentum-flow and stress distributions~\cite{Uji:2026zvn}. These developments establish a connection between small-$x$ gluon tomography and the mechanical properties of hadrons and nuclei encoded in matrix elements of the energy-momentum tensor.

Finally, we mention the closely related generalized parton distributions (GPDs), which provide the off-forward but transverse-momentum-integrated limit of the GTMD description and play a central role in nucleon tomography and spin physics. In particular, the second Mellin moments of GPDs enter the Ji sum rule, so that their small-$x$ behavior is relevant for determining the contribution of the low-$x$ region to the quark and gluon angular momentum. The small-$x$ behavior of GPDs can be studied directly in the high-energy/shockwave framework~\cite{Bhattacharya:2025fnz}. The small-$x$ evolution of the gluon helicity-flip GPD $E_g$ was studied in~\cite{Hatta:2022bxn}, where it was shown that, at zero skewness, $E_g$ exhibits BFKL-like evolution and saturation behavior similar to that of the unpolarized gluon GPD. The analysis has been extended to small but nonzero skewness, for both GPDs and GTMDs~\cite{Kovchegov:2025yyl,Kovchegov:2026gwb}. At leading logarithmic accuracy, and for $|x|\ll|\xi|\ll1$ or $|\xi|\ll|x|\ll1$, the skewness enters only through the rapidity argument of the dipole and odderon amplitudes, which becomes $Y=\ln\min\{1/|x|,1/|\xi|\}$~\cite{Kovchegov:2026gwb}. The real part of the elastic amplitude can in turn be obtained from the imaginary part of the dipole amplitude, through a more careful treatment of the initial condition of the non-linear evolution written in integral form. These two prescriptions could replace the phenomenological $R$-factors commonly used to account for skewness and for the real part of the amplitude in exclusive processes at small $x$, although they remain to be implemented in phenomenology. The same analysis shows that the unpolarized quark GPD receives an odderon contribution, confined to the ERBL region $|x|<|\xi|$.

\subsection{TMD factorization at small $x$ beyond leading power}
\label{sec:itmd}

As at moderate $x$, small-$x$ TMD factorization can also be studied beyond leading power in the ratio $k_\perp/Q$. In the small-$x$ regime, where the saturation scale $Q_s$ emerges as a characteristic scale of the target, it is useful to distinguish between kinematic power corrections, scaling as powers of $k_\perp/Q$, and saturation power corrections, involving ratios such as $Q_s/Q$, and to specify which are retained or neglected~\cite{Altinoluk:2019fui,Altinoluk:2019wyu}. For certain processes, an improved TMD (ITMD) factorization framework can be formulated in which kinematic power corrections $\mathcal{O}(k_\perp^n/Q^n)$ are resummed to all orders, while genuine saturation power corrections are neglected. This applies, for instance, to back-to-back dijet production in DIS and forward proton--nucleus collisions. The ITMD framework was first derived in~\cite{Kotko:2015ura,vanHameren:2016ftb} by identifying a common limit of the TMD limit of the CGC expressions for these processes~\cite{Dominguez:2010xd,Dominguez:2011wm} and of the high-energy factorization formula~\cite{Catani:1990eg,Catani:1994sq}.

The main advantage of this approach is that it extends the applicability of TMD factorization to small-$x$ cross sections beyond the back-to-back limit. By comparison, leading-power TMD factorization is generally employed when $k_\perp/Q$ is sufficiently small, typically below a value of order $0.2$--$0.3$, as in recent TMD extractions at moderate $x$~\cite{Bacchetta:2025ara,Moos:2025sal}.

In the ITMD framework, the hard factor is promoted from $H(Q)$ to $H(Q,k_\perp)$ and is computed from the matrix element for the hard partonic subprocess with an incoming off-shell gluon. It therefore depends on both the hard scale and the transverse momentum $k_\perp$. Crucially, however, the small-$x$ operator definition, including its full transverse-momentum dependence, remains unchanged. In the case of the WW gluon TMD, this relies on the possibility of expressing the product of two Wilson lines in terms of a transverse gauge link at $x^+\to\infty$~\cite{Boussarie:2020vzf}, such that
\begin{align}
    1-U(\boldsymbol{x})U^\dagger(\boldsymbol{y})
    &=\underbrace{ig\boldsymbol{r}^i
    \tilde{\boldsymbol{A}}_\perp^i(\boldsymbol{b})}_{\rm TMD}
    +\mathcal{O}(r_\perp^2) =ig\int_{\boldsymbol{y}_\perp}^{\boldsymbol{x}_\perp}
    \der\boldsymbol{z}\cdot
    \tilde{\boldsymbol{A}}_\perp(\boldsymbol{z})
    +\textrm{``genuine higher twist''},
\end{align}
where $\boldsymbol{r}=\boldsymbol{x}-\boldsymbol{y}$ and
$\boldsymbol{b}=(\boldsymbol{x}+\boldsymbol{y})/2$. The first equality, obtained by expanding in the transverse separation, introduces the transverse gauge field
$\tilde{A}_\perp^i(\boldsymbol{b})=(i/g)U(\boldsymbol{b})\partial^iU^\dagger(\boldsymbol{b})$
in the target light-cone gauge $\tilde A^-=0$. This is the standard TMD approximation used to connect a leading-order CGC calculation to the TMD framework. By contrast, the second equality retains the transverse gauge link and leads to the ITMD expression, which remains formulated in terms of the operator $U(\boldsymbol{b})\partial^iU^\dagger(\boldsymbol{b})$ entering the WW TMD defined in Eq.~\eqref{eq:WW-TMD}.

Consequently, the full dependence on $Q_s/k_\perp$ encoded in the transverse-momentum distribution is retained, whereas power corrections suppressed by $Q_s/Q$ are neglected. The connection to standard TMD factorization is particularly simple: the latter is recovered by setting $k_\perp=0_\perp$ in the hard factor. A comparison of TMD, ITMD, and full CGC calculations for inclusive back-to-back dijet production at leading order can be found in~\cite{Boussarie:2021ybe}. This study explicitly demonstrates that, in the limit $Q_s\ll Q$, the leading-order CGC cross section approaches the ITMD result for all values of $k_\perp$, not merely in the back-to-back regime $k_\perp\ll Q$.

At leading order in the CGC, the ITMD formalism has been applied to the phenomenology of inclusive particle production in forward proton--nucleus collisions~\cite{Bury:2020ndc,vanHameren:2023oiq,Ganguli:2023joy,Al-Mashad:2022zbq,Caucal:2025zkl} and DIS~\cite{vanHameren:2021sqc,Altinoluk:2021ygv}. An important challenge for future work is to establish the structure of ITMD factorization beyond leading order. Its extension beyond the eikonal approximation may also be nontrivial. In an operator basis that interpolates between the eikonal and the back-to-back limits of dijet production in DIS, the resummation of transverse kinematic corrections generates a family of longitudinal phases. These phases become indistinguishable in the eikonal limit, where the ITMD structure is recovered, but at nonzero $x$ they cannot be absorbed into a single conventional TMD operator~\cite{Kar:2026cnx}.

Regarding genuine saturation power corrections, the connection between the CGC framework, which formally resums contributions of all twists in the strict $x\to 0$ limit, and the higher-twist formalism, which incorporates finite-$x$ corrections while retaining the first subleading terms in the expansion in inverse powers of the hard scale, has been explored in~\cite{Fu:2023jqv,Fu:2024sba} for inclusive photon production in proton--nucleus collisions. It would be interesting to extend this analysis to less inclusive observables, such as two-particle correlations, to identify and control the leading corrections in the $Q_s/Q$ expansion implicit in both the TMD and ITMD frameworks.

\subsection{Monte Carlo simulations and heavy-ion collisions}

Monte Carlo event generators have become standard tools for simulating particle production in high-energy collisions. Compared with analytic approaches, their main advantages are their versatility in the range of processes they can describe and their ability to incorporate hadronization models, such as those implemented in \textsc{Pythia}~\cite{Bierlich:2022pfr} or \textsc{Herwig}~\cite{Bahr:2008pv}. They can therefore provide a more realistic description of complete experimental events. For this reason, they are also essential tools for experimentalists seeking to understand detector responses and assess the systematic uncertainties associated with their measurements. In view of the future EIC, it will be important to develop Monte Carlo generators that incorporate intrinsic $k_\perp$ effects for both gluons and quarks in the target, together with saturation effects and non-linear small-$x$ evolution in the kinematic regime where these effects become relevant.

We can mention several works going in this direction. First, in the dilute regime $k_\perp \gg Q_s$, one possible approach is to use the CCFM equation~\cite{Ciafaloni:1987ur,Catani:1989sg,Marchesini:1994wr} or the Parton Branching Method~\cite{Hautmann:2017xtx,Hautmann:2017fcj,BermudezMartinez:2018fsv}. These evolution equations incorporate Sudakov factors that resum the large logarithms of the ratio $k_\perp/Q$ discussed in Sec.~\ref{subsub:Sudakovresum}. At small $x$, they reduce to BFKL evolution and therefore resum the large $\ln(1/x)$ logarithms in the dilute regime. In this sense, they provide a branching-based formulation of TMD evolution that captures both the Sudakov and small-$x$ logarithms. The advantage of the Parton Branching approach over the CCFM equation is that it reproduces the correct collinear limit, including the full DGLAP splitting functions, while also accounting for quark contributions to the parton densities. It therefore provides a smooth connection between the collinear and small-$x$ regimes for dilute parton densities, but does not account for non-linear effects associated with high parton densities.

A major advantage of the CCFM and Parton Branching approaches is their parton-shower interpretation, which makes them particularly well suited for Monte Carlo implementation. Saturation effects can be effectively incorporated into the initial condition for the intrinsic $k_\perp$ distribution of gluons in the target. However, this does not account for the non-linear nature of small-$x$ evolution itself: the subsequent evolution remains linear and, in the small-$x$ limit, is governed by the BFKL equation. Consequently, gluon recombination and the dynamical generation of saturation are not included in CCFM and Parton Branching Method based showers. A Monte Carlo implementation of the Parton Branching Method has been developed in the generator CASCADE~\cite{CASCADE:2010clj,CASCADE:2021bxe}. The Parton Branching Method has also been generalized~\cite{Hautmann:2022xuc} to include transverse-momentum dependent splitting functions~\cite{Hautmann:2012sh,Hentschinski:2017ayz,Gituliar:2015agu,Hentschinski:2016wya,Hentschinski:2021lsh}, and implemented in a parton shower~\cite{Keersmaekers:2021arn}.

Saturation has entered event generators in two distinct ways. In the first, the non-linear evolution is computed beforehand, in a BK-evolved dipole amplitude or a small-$x$ TMD, and sampled in the hard process, while radiation from the target is either absent or generated by a linear shower. The MC-CGC generator samples the hybrid or $k_\perp$-factorized cross section for forward hadron production in $pp$ collisions with running-coupling BK dipoles, and adds radiation from the projectile, multiple scatterings and string fragmentation~\cite{Deng:2014vda,Fujii:2026ccu}. Radiation from the dense target is not generated as an exclusive shower in this approach, but is absorbed into the evolved dipole. Parton-level generators with off-shell initial states, such as KaTie~\cite{vanHameren:2016kkz}, implement high-energy and ITMD factorization with saturation entering through the input gluon distributions. Back-to-back dijets in DIS have been generated from the WW distribution and its linearly polarized partner, with showering and hadronization delegated to \textsc{Pythia}~\cite{Dumitru:2018kuw}. The CGC-py generator couples the full leading-order CGC cross section for $\gamma^*A\to q\bar q X$, without a correlation-limit expansion, to a backward initial-state shower of the Parton Branching type guided by the CSS-evolved WW distribution, and to \textsc{Pythia} hadronization~\cite{Duan:2026jrk}. It describes the H1 charged-hadron spectra in the current region, and finds that collinear showering and small-$x$ evolution contribute comparably to the decorrelation of dihadrons, with nuclear effects that remain modest at EIC energies.

In the second, the non-linear evolution is generated inside the shower itself. A backward initial-state shower has been constructed from the Gribov--Levin--Ryskin (GLR) equation~\cite{Gribov:1984tu}, in which recombination enters through the $2\to1$ fusion of gluon ladders and appears as an additional loss in the no-branching probability~\cite{Shi:2022hee}. The GLR equation is used rather than the BK equation because the momentum-space dipole amplitude has no probabilistic interpretation, and because keeping the emitted gluons exclusive would require the higher-point correlators of the JIMWLK hierarchy. The algorithm was extended with the kinematic constraint and with a Collins--Soper stage that generates soft emissions down to a scale of order $Q_s$ before handing over to the GLR stage, which provides a first Monte Carlo realization of the joint resummation of Sudakov and small-$x$ logarithms discussed in Sec.~\ref{sec:sudakov}~\cite{Shi:2023ejp}. These showers have been validated against numerical solutions of the underlying equations rather than against data. The GLR shower has since been embedded in the eHIJING generator for $eA$ collisions~\cite{Ke:2023xeo}, matched to a collinear shower above $x=0.01$, and completed with beam remnants, final-state radiation and hadronization~\cite{Ke:2026ytl}. In this framework, the parton-level saturation broadening appears to be strongly diluted in the nucleon energy correlator and in the dihadron decorrelation, while the global hadronic recoil retains it (Sec.~\ref{sec:recoil}). A shower consistent with BK or JIMWLK evolution, including multiple rescattering in the dense target and the linearly polarized gluon distributions, remains to be constructed.

\paragraph{Heavy-ion collisions}

This review has mostly focused on dilute-dense collisions, for which high-energy factorization at small $x$ has been established at fixed order for a number of processes. Although its theoretical status in dense-dense collisions is not yet formally established, the framework can nevertheless be phenomenologically extended to heavy-ion collisions to provide a baseline description of particle production, including saturation effects but neglecting final-state interactions. In particular, this baseline does not include energy-loss effects or other modifications of the produced particles due to their subsequent interaction with the quark-gluon plasma. As usual in TMD factorization, the processes which can be described using this framework are those with a hard scale $Q$ and a semi-hard one $k_\perp\ll Q$: this includes two-particle correlations in heavy-ion collisions for instance. Phenomenological studies of two-particle correlations in heavy-ion collisions with a TMD factorized cross-section as a baseline for jet quenching effects can be found in~\cite{Mueller:2016gko} for dijet production and~\cite{Adhya:2024pgm} for $\gamma$-jet correlations. Unintegrated parton distributions and nuclear TMDs derived from the CGC framework can also provide saturation-based descriptions of the initial state of heavy-ion collisions. This approach has been implemented in the McDIPPER Monte Carlo generator~\cite{Garcia-Montero:2023gex} to produce event-by-event profiles of the initial energy and charge densities.

\section{Challenges and prospects}

\label{sec:outlook}

In this review, we have surveyed the status of transverse-momentum-dependent factorization at small $x$, with a particular emphasis on the dense QCD regime, where the intrinsic transverse momentum of partons inside the target is of the order of the saturation scale $Q_s$. Probing saturation in processes that admit TMD factorization, as well as understanding the quantum many-body dynamics of QCD in dense nuclei, is one of the main motivations for studying the small-$x$ behavior of TMDs. By comparison, the standard TMD factorization framework at moderate $x$ has reached a more advanced stage in the determination of its non-perturbative inputs from data~\cite{Bacchetta:2025ara,Moos:2025sal}, both for the intrinsic $k_\perp$ distribution of partons and for the non-perturbative component of the Collins--Soper kernel~\cite{Avkhadiev:2024mgd}.

Since the early 2010s, however, the small-$x$ TMD field has seen important developments on both the conceptual and the phenomenological fronts. At leading order, the correspondence between TMDs and Wilson-line correlators has been established for a broad class of inclusive processes~\cite{Dominguez:2010xd,Dominguez:2011wm}, and at large $N_c$ the resulting gluon and sea-quark distributions reduce to a small number of building blocks~\cite{Marquet:2016cgx,Caucal:2025xxh}. Diffractive and target-fragmentation processes have been brought into the same framework~\cite{Iancu:2021rup,Hatta:2022lzj,Hauksson:2024bvv,Caucal:2025qjg}. At next-to-leading order, back-to-back dijet production in DIS has been shown to factorize into the WW distribution, a soft factor and a coefficient function, once the gluon phase space is shared consistently between high-energy and Sudakov resummation~\cite{Taels:2022tza,Caucal:2022ulg,Caucal:2023nci}. Beyond the eikonal approximation, small-$x$ evolution has been derived for the helicity distributions and for the full set of leading-twist quark TMDs~\cite{Cougoulic:2022gbk,Santiago:2023rfl,Adamiak:2024khm}, and it now enters global extractions of the helicity distributions~\cite{Adamiak:2023yhz}. On the phenomenological side, calculations that combine small-$x$ evolution, Sudakov resummation and kinematic power corrections are now confronted with RHIC data~\cite{Caucal:2025zkl}.

While these results place small-$x$ TMDs on a firmer footing, several conceptual questions remain open. First, factorization at small $x$ is established at fixed order and for particular processes, and no all-order argument comparable to that available at moderate $x$ exists. Establishing high-energy factorization at next-to-leading-logarithmic accuracy in the dense regime requires two-loop calculations, either of the operators themselves or of observables that factorize onto them (Sec.~\ref{sec:jimwlk}). Establishing a consistent TMD factorization framework at two-loop accuracy is an important challenge for the coming decade. Together with the recent calculations of the three-loop BK equation in the planar limit~\cite{Caron-Huot:2016tzz,Brunello:2025rhh}, this would pave the way for a new era of precision phenomenology in the field. Second, for processes with colored partons in both the initial and the final state, such as forward dihadron production in $pA$ collisions, it is not known whether the finite operator basis found at leading order survives at next-to-leading order, nor whether CSS resummation applied to each distribution can be matched to a next-to-leading-order small-$x$ calculation (Secs.~\ref{sec:tmds} and~\ref{subsub:Sudakovresum}). Even for the dipole gluon distribution the analogue of the CSS equation has been established only at double-logarithmic accuracy, and the single-logarithmic Sudakov terms beyond the coefficient $B$ may not exponentiate in the standard way. Third, the non-perturbative input remains model dependent. The initial conditions of the BK and JIMWLK equations are fitted to inclusive data, the Gaussian truncation of the higher-point correlators, which reproduces the TMDs at the initial condition, underestimates the speed of JIMWLK evolution for most of the color structures that enter them~\cite{Cougoulic:2026drr,Cougoulic:2026sww}, and the initial conditions for the polarized Wilson-line operators and for the spin-dependent odderon are even less constrained. Since lattice QCD is unlikely to reach this regime in the near future, these inputs will have to be constrained by data.

A second set of challenges concerns the connection with the moderate-$x$ regime. Sub-eikonal corrections provide a systematic route toward it, but the modeling of the correlators of the quark background field and their high-energy evolution are only beginning to be addressed (Sec.~\ref{sec:beyond-eikonal}). In the spin sector, the asymptotics of the quark TMDs are known at large $N_c$ in the linearized double-logarithmic approximation, whereas the gluon $g_{1T}$ has not been computed, a helicity-dependent generalization of the JIMWLK equation is not yet known, and the matching onto CSS evolution has not been studied. From the opposite direction, the ITMD framework resums the kinematic power corrections in $k_\perp/Q$ at leading order, and its structure beyond leading order, together with the genuine saturation corrections in $Q_s/Q$, remains to be understood (Sec.~\ref{sec:advanced}). A unified description of TMDs across the full range of $x$ will require these lines of work to meet.

On the phenomenological side, the main lesson of the past decade is that no single observable establishes saturation. While the suppression of the away-side peak in forward dihadron production is seen at RHIC, Sudakov radiation dominates the dijet imbalance in EIC kinematics, collinear nuclear distributions can produce suppressions of similar size in some observables, and hadronization may dilute the parton-level signal (Sec.~\ref{sec:pheno}). Evidence for saturation will therefore rest on the dependence of nuclear ratios on $x$, on nuclear size and on the hard scale, across several observables described within the same framework and at the same accuracy. This requires next-to-leading-order calculations for processes that are so far known only at leading order, such as forward photon--jet production in $pA$ collisions and the energy correlators, global analyses that use the same non-perturbative input for all channels, and Monte Carlo event generators in which the non-linear evolution of the target is consistent with the BK or JIMWLK equations.

The coming decade offers a favorable situation for this program. The EIC will measure semi-inclusive, dijet, dihadron and diffractive observables in electron--proton and electron--nucleus collisions over a range of nuclear sizes, with polarized beams that constrain the small-$x$ contribution to the proton spin~\cite{AbdulKhalek:2021gbh}, while forward detectors at the LHC, such as the ALICE FoCal, will extend the reach of $pA$ measurements toward smaller $x$. We hope that the developments reviewed here, together with these measurements, will allow the saturation regime to be studied with TMD observables at a level of precision comparable to that reached by TMD phenomenology at moderate $x$.

\begin{ack}[Acknowledgments]
We are grateful to Gabriel Santiago for a careful reading of Sec.~\ref{sec:beyond-eikonal} and for helpful comments. P.~C. is funded by the Agence Nationale de la Recherche under
grant ANR-25-CE31-5230 (TMD-SAT) and by the IN2P3 Master Project QHAT. F.~S. is supported by the Laboratory Directed Research and Development of Brookhaven National Laboratory and RIKEN-BNL Research Center, as well as the National Science Foundation (NSF) within the framework of the JETSCAPE collaboration, under grant number OAC-2514008 (CSSI:C-SCAPE). P.~C. and F.~S. acknowledge the Saturated Glue (SURGE) Topical Theory Collaboration, funded by the U.S. Department of Energy, Office of Science, Office of Nuclear Physics.
\end{ack}

{\small\setstretch{1.0}\bibliographystyle{elsarticle-num}
\bibliography{refs.bib}}

\end{document}